\documentclass[fleqn,usenatbib]{mnras}
\usepackage{newtxtext,newtxmath}
\usepackage[T1]{fontenc}

\DeclareRobustCommand{\VAN}[3]{#2}
\let\VANthebibliography\thebibliography
\def\thebibliography{\DeclareRobustCommand{\VAN}[3]{##3}\VANthebibliography}

\usepackage{graphicx}	
\usepackage{amsmath}	
\usepackage{multirow}
\usepackage{lipsum}
\usepackage{wasysym}
\usepackage{color}
\usepackage{arydshln}
\usepackage{tablefootnote}

\newcommand{\aum}{AU\,Mic}
\newcommand{\pr}{$P_{\rm{rot}}$}
\newcommand{\vs}{$v \sin i_{\rm{rot}}$}
\newcommand{\kms}{km\,s$^{-1}$}
\newcommand{\ms}{m\,s$^{-1}$}

\newcommand{\msun}{M$_{\odot}$}
\newcommand{\rsun}{R$_{\odot}$}
\newcommand{\rs}{$R_{\rm{S}}$}
\newcommand{\bl}{B$_{\ell}$}

\newcommand{\crr}{$\chi_{\rm{r}}^{2}$}
\newcommand{\chs}{$\chi^{2}$}

\newcommand{\mdw}{M\,dwarf}

\newcommand{\teff}{$T_{\rm{eff}}$}
\newcommand{\lgg}{$\log g$}

\newcommand{\voo}{$V_{0}$}
\newcommand{\vr}{$\boldsymbol{V_{\rm{r}}}$}
\newcommand{\vp}{$\boldsymbol{V_{\rm{p}}}$}
\newcommand{\vj}{$\boldsymbol{V_{\rm{j}}}$}
\newcommand{\kp}{$K_{\rm{p}}$}
\newcommand{\mpp}{$M_{\rm{p}}$}
\newcommand{\mpear}{$M_{\oplus}$}
\newcommand{\php}{$\phi_{\rm{p}}$}
\newcommand{\porb}{$P_{\rm{orb}}$}
\newcommand{\tze}{$T_{0}$}
\newcommand{\hypv}{$\boldsymbol{\theta}$}

\newcommand{\oeq}{$\Omega_{\rm{eq}}$}
\newcommand{\dome}{d$\Omega$}

\defcitealias{plavchan2020}{P20}

\title[SPIRou observations of AU Mic]{Investigating the young AU~Mic system with SPIRou: large-scale stellar magnetic field and close-in planet mass}

\author[B. Klein et al.]{
 \parbox[h]{\textwidth}{
Baptiste Klein$^{1}$\thanks{E-mail: baptiste.klein@irap.omp.eu},
Jean-Fran\c{c}ois Donati$^{1}$,
Claire Moutou$^{1}$,
Xavier Delfosse$^{2}$,
Xavier Bonfils$^{2}$,
Eder Martioli$^{3,4}$,
Pascal Fouqu\'e$^{1,5}$,
Ryan Cloutier$^{6}$,
\'Etienne Artigau$^{7}$,
René Doyon$^{7}$,
Guillaume H\'ebrard$^{3}$,
Julien Morin$^{8}$,
Julien Rameau$^{2}$,
Peter Plavchan$^{9}$,
Eric Gaidos$^{10}$
}
\vspace{6pt}\\
$^{1}$Universit\'e de Toulouse, CNRS, IRAP, 14 av. Belin, 31400 Toulouse, France\\
$^{2}$CNRS, IPAG, Universit\'e Grenoble Alpes, 38000 Grenoble, France\\
$^{3}$Institut dAstrophysique de Paris, UMR7095 CNRS, Universit\'e Pierre \& Marie Curie, 98bis boulevard Arago, 75014 Paris, France\\
$^{4}$Laboratorio Nacional de Astrofisica (LNA/MCTI), Rua Estados Unidos, 154, Itajuba, MG, Brazil\\
$^{5}$CFHT Corporation; 65-1238 Mamalahoa Hwy; Kamuela, Hawaii 96743; USA\\
$^{6}$Center for Astrophysics | Harvard \& Smithsonian, 60 Garden Street, Cambridge, MA, 02138, USA\\
$^{7}$Institut de Recherche sur les Exoplan\`etes (IREx), D\'epartement de Physique, Universit\'e de Montr\'eal, C.P. 6128, Succ. Centre-Ville, Montr\'eal, QC, H3C 3J7,Canada\\
$^{8}$LUPM, Universit\'e de Montpellier, CNRS, Place Eug\`ene Bataillon, F-34095 Montpellier, France\\
$^{9}$Department of Physics and Astronomy, George Mason University, Fairfax, VA, 22030, USA\\
$^{10}$Department of Earth Sciences, University of Hawai'i at Manoa, Honoluu, HI 96822 USA
}

\date{Accepted. Received in original form 2020 November 24}

\pubyear{2020}

\begin{document}
\label{firstpage}
\pagerange{\pageref{firstpage}--\pageref{lastpage}}
\maketitle

\begin{abstract}
We present a velocimetric and spectropolarimetric analysis of 27 observations of the 22-Myr M1 star AU Microscopii (\aum) collected with the high-resolution \textit{YJHK} (0.98-2.35\,$\mu$m) spectropolarimeter SPIRou from 2019 September 18 to November 14. Our radial velocity (RV) time-series exhibits activity-induced fluctuations of 45\,\ms\ RMS, $\sim$3$\times$ smaller than those measured in the optical domain, that we filter using Gaussian Process Regression. We report a  3.9$\sigma$-detection of the recently-discovered 8.46\,d-transiting planet \aum~b, with an estimated mass of 17.1$^{+4.7}_{-4.5}$\,\mpear\ and a bulk density of 1.3\,$\pm$\,0.4\,g\,cm$^{-3}$, inducing a RV signature of semi-amplitude K=8.5$^{+2.3}_{-2.2}$\,\ms\ in the spectrum of its host star. A consistent detection is independently obtained when we simultaneously image stellar surface inhomogeneities and estimate the planet parameters with Zeeman-Doppler Imaging (ZDI). Using ZDI, we invert the time series of unpolarized and circularly-polarized spectra into surface brightness and large-scale magnetic maps. We find a mainly poloidal and axisymmetric field of 475\,G, featuring, in particular, a dipole of 450\,G tilted at 19\degr\ to the rotation axis. Moreover, we detect a strong differential rotation of \dome\,=\,0.167\,$\pm$\,0.009\,rad/d shearing the large-scale field, about twice stronger than that shearing the brightness distribution, suggesting that both observables probe different layers of the convective zone. Even though we caution that more RV measurements are needed to accurately pin down the planet mass, \aum~b already appears as a prime target for constraining planet formation models, studying the interactions with the surrounding debris disk, and characterizing its atmosphere with upcoming space- and ground-based missions.
\end{abstract}




\section{Introduction}

Close-in planetary systems orbiting and transiting pre-main-sequence (PMS) stars are key targets to improve our understanding of how planets form and evolve. Their orbital parameters (e.g., orbit ellipticity and spin-orbit obliquity) and the composition of their atmosphere can yield essential information about their formation history \citep{baruteau2016,madhusudhan2019}. Moreover, the evolution of their bulk density during the early stages of their lives is critically needed to constrain planet formation and evolution models \citep[e.g.][]{alibert2005,mordasini2012a,mordasini2012b}. This requires to precisely measure both planet masses, by monitoring the radial velocity (RV) of their host star, and radii, through the relative depth of their photometric transit curve.

PMS stars, however, exhibit intense magnetic activity whose underlying dynamo processes are not fully-understood yet \citep[e.g.,][]{donati2014,yu2019,hill2019}. This activity generates large bright and dark features at the surface of the star, leading to fluctuations in both photometric and RV curves that are much stronger than planet signatures, rendering them extremely difficult to detect. As a consequence, only a few close-in newborn planets have been unveiled so far, none of them having a well-constrained bulk density \citep[see][only two of these systems are transiting]{donati2016,johns_krull2016,david2016,mann2016,yu2017,david2019,david2019b}.

AU Microscopii (\aum, GJ\,803, HD\,197481, HIP\,102409) is an active nearby ($d$\,=\,9.72\,pc) M1 PMS star located in the $\sim$22\,Myr-old $\beta$~Pic moving group \citep{mamajek2014,malo2014,messina2016}. The star has been widely studied over the past decades for its intense magnetic activity generating numerous phenomena, e.g., $\sim$0.1\,mag photometric variations in the visible domain \citep[e.g.,][]{torres1973,rodono1986}. \aum\ hosts a resolved edge-on debris disk at distances ranging from $\sim$35 to 210\,au, offering an excellent framework to study planet formation and evolution \citep[e.g.][]{kalas2004,wilner2012,macgregor2013,matthews2015,daley2019}. Asymmetric fast moving structures have been discovered in \aum's debris disk \citep{boccaletti2015,boccaletti2018,daley2019}. The origin of these features, tentatively explained by interactions between the stellar wind and the disk \citep{chiang2017}, or the presence of a massive body in the inner region of the disk \citep{sezestre2017}, remains unclear. A close-in transiting Neptune-sized planet was newly detected around \aum\ from photometric observations collected with TESS and Spitzer space missions \citep[][hereafter \citetalias{plavchan2020}]{plavchan2020}. Given the brightness of the star and the relatively small semi-major axis of the planet orbit (0.066\,au), \aum\,b stands as the best transiting PMS target for a velocimetric mass measurement. However, RV time-series measured from spectra covering the optical wavelength domain exhibit large activity-induced fluctuations of 115 to 175\,\ms\ RMS, leading \citetalias{plavchan2020} to report no more than an upper limit of 3.4\,$M_{\neptune}$ (58.3\,\mpear) for the mass of \aum~b (where $M_{\neptune}$ denotes the mass of Neptune).


The near-infrared (nIR) spectropolarimeter SPIRou at the Canada-France-Hawaii telescope \citep[CFHT;][]{donati2018,donati2020} is the ideal instrument to detect \aum~b and measure its mass. Its high resolving power of $\sim$70\,000 throughout the \textit{YJHK} spectral bands (i.e., 0.98-2.35\,nm) makes it well-suited to provide precise RVs for as bright a star as \aum\ \citep[H\,=\,4.831;][]{cutri2003}. Moreover, the RV signal generated by stellar activity for PMS stars is expected to be significantly weaker in the nIR than in the optical domain \citep{mahmud2011,crockett2012,klein2020}, making it easier to separate from the planet signature. Thanks to its spectropolarimetric capabilities, SPIRou has the additional potential to detect and reconstruct the surface topology of the large-scale magnetic field of \aum\ \citep{donati2009}, allowing one not only to filter RV curves from the modeled activity, but also to constrain the underlying dynamo processes.

In this study, we present velocimetric and spectropolarimetric data of \aum\ consisting of 27 unpolarized and circularly-polarized spectra collected with SPIRou from 2019 September 18 to November 14. The observations and the data reduction are described in Sec.~\ref{sec:data_red}. In Sec.~\ref{sec:rv_an}, we detail the RV measurement process as well as the detection of a planet signature from our RV time-series. In Sec.~\ref{sec:bright_im}, we present independent methods based on Zeeman-Doppler imaging to confirm the planet signature while filtering the stellar activity. The inversion of unpolarized and circularly-polarized profiles into brightness and large-scale magnetic maps is detailed in Sec.~\ref{sec:mag_an}. Finally, we investigate the chromospheric activity of the star and its correlation with the magnetic map and RV time-series in Sec.~\ref{sec:act_ind}, before discussing, in Sec.~\ref{sec:conclusion}, the impact of these results on our understanding of star/planet formation.


\section{Observations and data reduction}\label{sec:data_red}

\aum\ was observed between 2019 September 18 and November 14 using the nIR spectropolarimeter and high-precision velocimeter SPIRou at the CFHT atop MaunaKea, Hawaii \citep[][]{donati2018,donati2020}. Our data set consists of 27 spectropolarimetric observations collected at a rate of one polarization sequence (i.e., four individual exposures taken at different orientations of the polarimeter retarders) per night during CFHT bright time periods. All observations were obtained in average seeing conditions (median seeing of $\sim$1\arcsec) and under a median airmass of 1.6. The full journal of observations is given in Tab.~\ref{tab:list_obs}.

Data reduction is carried out using a version of the ESPaDOnS and NARVAL reduction pipeline \citep[Libre-ESpRIT,][]{donati1997a}, adapted for SPIRou observations \citep[see][]{donati2020}. The subexposures within each polarization sequence are individually extracted using the method described in \citet{horne1986}, and corrected for the blaze function, estimated from flat-field exposures collected prior to the observations. The individual subexposures are then combined into unpolarized (Stokes $I$) and circularly-polarized (Stokes $V$) spectra, in a way to remove systematics in circular polarization spectra at first order \citep{donati1997a,bagnulo2009}. Telluric lines from the Earth's atmosphere are modeled and filtered from our Stokes $I$ spectra with the method detailed in \citet{artigau2014}. The reduced intensity spectra exhibit peak signal-to-noise ratios (S/N) per pixel (2.28\,\kms\ velocity bin) ranging from 548 to 753 with a median value of 678 (see Tab.~\ref{tab:list_obs}).



We apply Least-Squares Deconvolution \citep[LSD,][]{donati1997a} to extract average Stokes $I$ and $V$ profiles from our set of reduced spectra. For RV measurement and brightness reconstruction (see Sec~\ref{sec:rv_an} and~\ref{sec:bright_im}), we use a weighted mask of $\approx$3000 atomic and molecular lines, empirically built from SPIRou observations of the M2V star Gl\,15A. For the magnetic analysis (see Sec~\ref{sec:mag_an}), we use a mask of atomic lines computed from an \textsc{atlas9} local thermodynamical equilibrium model \citep{kurucz1993}, assuming an effective temperature \teff\,=\,3750\,K and a surface gravity \lgg\,=\,4.5 (see Tab.~\ref{tab:star_prop}). The final mask spans the entire SPIRou spectral range (i.e., $YJHK$ bands), and contains about 3600 Zeeman-sensitive atomic lines with known Land\'e factor and depth relative to the continuum down to 1\%. The extracted Stokes $I$ and $V$ LSD profiles feature a mean central wavelength of 1700\,nm, an effective Land\'e factor of 1.2, and a depth of 0.24 with respect to the continuum. As a result of Zeeman broadening affecting atomic lines much more than molecular lines, the Stokes $I$ LSD profiles extracted from the atomic line list are $\sim$2$\times$ broader than those extracted from the empirical mask. The wavelengths of each LSD profile are finally corrected from the Barycentric Earth RV. A similar data reduction process applied to SPIRou velocimetric observations of the inactive star Gl\,699 (collected at almost the same epochs as those of \aum), yields a median absolute deviation about the mean of 3\,\ms\ and a standard deviation of $\sim$5\,\ms. We thus take 5\,\ms\ as a conservative error bar for our RV measurements of \aum\ (including a photon noise $\sigma_{\rm{ph}}$ that scales with the inverse S/N and contributes from 2.1 to 2.9\,\ms\ with a median of 2.2~\ms\ to the error budget).





Clear Zeeman signatures of full-amplitude up to 0.4\% with respect to the unpolarized continuum are detected in the Stokes $V$ LSD profiles (see Sec.~\ref{sec:mag_an}), in good agreement with previous spectropolarimetric measurements of \aum\ in the optical domain \citep{berdyugina2006}. For each observation, we compute the longitudinal magnetic field, \bl, using the relation introduced in \citet{donati1997a}. The resulting \bl\ values, listed in tab.~\ref{tab:list_obs}, range from -48 to 83\,G with a median of 33\,G, and a typical 1$\sigma$ uncertainty of 3.2\,G. The analysis of the \bl\ time-series is described in Sec.~\ref{sec:act_ind}.

In what follows, the rotation cycle of the star is computed from the stellar rotation period \pr\,=\,4.86\,d \citepalias{plavchan2020}, with the reference time BJD\,=\,2458651.993 (called \tze\ in the following), corresponding to the mid-transit of \aum~b observed with SPIRou on June 17, 2019 \citep[see the dedicated analysis in][]{martioli2020}.


\setlength{\tabcolsep}{4.pt}

\begin{table}
    \centering
    \caption{Journal of spectropolarimetric observations of \aum\ with SPIRou. All polarization sequences consist of 4 individual exposures of 195\,s each. The first column lists the BJDs at mid-exposure (the first and last observations corresponding respectively to September 18 and November 14). In columns~2, 3 and~4, we respectively give the peak S/N (per pixel, i.e., 2.28~\kms\ velocity bin) and the RMS noise level in Stokes $I$ and Stokes $V$ LSD profiles (resp. called $\sigma_{\rm{I}}$ and $\sigma_{\rm{V}}$; note that $\sigma_{\rm{I}}$ is given for the Stokes $I$ extracted with the empirical line mask and empirically accounts for systematics in the line profiles; see Sec.~\ref{sec:bright_im}), with respect to the unpolarized continuum, $I_{\rm{C}}$. The longitudinal magnetic field, estimated from our sequences of spectra, is given in Column~5. In column~6, we list the RVs measured by fitting a Gaussian function to the Stokes $I$ LSD profiles. We adopt a conservative 1$\sigma$-uncertainty of 5\,\ms\ on the RVs. Finally, the rotational cycles listed in column~7 are computed using the reference time BJD\,$=$\,2458651.993 (corresponding to the mid transit of \aum~b on 2019 June 19) and a rotation period of 4.86\,d.}
    \label{tab:list_obs}
\begin{tabular}{ccccr@{$\pm$}lcc}
\hline
 BJD   &  S/N & $\sigma_{\rm{I}}$ & $\sigma_{\rm{V}}$ & \multicolumn2c{\bl} & RV & Cycle \\
 $[2457000+]$  &   & [$10^{-4} I_{\rm{C}}$]  & [$10^{-4} I_{\rm{C}}$] & \multicolumn2c{[G]} & [\ms] & \\
 \hline
1744.8212 & 708 & 4.10 & 1.41 & -43.9\, & \,3.1 & 59.5 & 0.10\\
1750.7542 & 740 & 4.27 & 1.60 & 40.6\, & \,3.7 & -18.2 & 1.32\\
1751.7453 & 753 & 4.07 & 1.47 & 31.3\, & \,3.2 & -52.4 & 1.53\\
1752.7898 & 740 & 4.11 & 1.39 & 48.6\, & \,2.9 & 27.9 & 1.74\\
1758.7288 & 678 & 4.13 & 1.36 & 16.2\, & \,2.8 & 51.3 & 2.96\\
1759.8053 & 617 & 4.10 & 1.44 & 15.9\, & \,3.2 & 69.2 & 3.18\\
1760.7278 & 549 & 4.10 & 1.67 & 48.5\, & \,3.7 & -29.6 & 3.37\\
1761.7305 & 736 & 4.18 & 1.40 & 24.8\, & \,2.9 & -87.2 & 3.58\\
1762.7315 & 724 & 4.09 & 1.31 & 52.6\, & \,2.7 & -2.9 & 3.79\\
1764.7571 & 734 & 4.10 & 1.75 & 23.6\, & \,3.9 & 34.5 & 4.20\\
1765.7694 & 739 & 3.98 & 1.44 & 34.4\, & \,3.2 & -39.8 & 4.41\\
1769.7438 & 742 & 4.13 & 1.41 & 32.6\, & \,3.1 & 22.1 & 5.23\\
1770.7407 & 678 & 4.07 & 1.56 & 33.4\, & \,3.4 & -41.6 & 5.43\\
1771.7212 & 664 & 4.04 & 1.45 & 16.4\, & \,3.1 & -55.8 & 5.64\\
1772.7416 & 629 & 4.17 & 1.57 & 47.2\, & \,3.4 & 19.4 & 5.85\\
1787.7155 & 548 & 4.07 & 1.54 & -4.7\, & \,3.2 & 28.3 & 8.93\\
1788.7045 & 569 & 4.04 & 1.49 & 4.2\, & \,3.3 & 41.3 & 9.13\\
1789.7367 & 623 & 4.02 & 1.38 & 82.5\, & \,3.2 & -32.6 & 9.34\\
1790.7010 & 643 & 4.08 & 1.66 & 12.4\, & \,3.6 & -90.1 & 9.54\\
1791.6983 & 639 & 4.05 & 1.35 & 40.3\, & \,2.8 & 2.1 & 9.75\\
1792.6976 & 667 & 4.04 & 1.50 & -24.3\, & \,3.1 & 19.1 & 9.95\\
1796.6859 & 569 & 4.03 & 1.73 & 43.7\, & \,3.6 & -13.5 & 10.77\\
1797.7318 & 567 & 4.35 & 1.70 & -48.0\, & \,3.7 & 19.7 & 10.98\\
1798.6873 & 609 & 4.03 & 1.53 & 69.5\, & \,3.5 & 27.0 & 11.18\\
1799.6883 & 559 & 4.11 & 1.68 & 70.0\, & \,3.9 & -10.3 & 11.39\\
1800.6896 & 724 & 4.06 & 1.40 & 8.8\, & \,3.0 & -46.2 & 11.60\\
1801.6873 & 703 & 4.12 & 1.55 & 44.3\, & \,3.2 & 0.0 & 11.80\\
\hline
\end{tabular}
\end{table}

\begin{table}
    \centering
    \caption{Stellar properties of \aum\ used in this study. In particular, the stellar inclination, $i_{\rm{rot}}$, is assumed equal to the inclination $i_{\rm{orb}}$ of the planet orbit reported in \citetalias{plavchan2020} and consistent with the rotation-orbit alignment recently reported \citep[see][]{martioli2020,palle2020,hirano2020}. Note that ZDI reconstructions are performed assuming a stellar inclination of $i_{\rm{ZDI}}$\,=\,80\degr\ (see Sec.~\ref{sec:bright_im}). Note that our estimate of \vs\ is consistent with the value reported in \citet{scholz2007} and \citet{weise2010}.}
    \begin{tabular}{ccc}
    \hline
    Parameter & Value & Reference \\
    \hline
    Distance &  9.7248\,$\pm$\,0.0046\,pc & \cite{gaia2018}\\
  \teff & 3700\,$\pm$\,100\,K & \cite{afram2019} \\
  Radius (\rs) & 0.75\,$\pm$\,0.03\,\rsun &  \citetalias{plavchan2020} \\
  Mass ($M_{\rm{S}}$) & 0.50\,$\pm$\,0.03\,\msun & \citetalias{plavchan2020} \\
  $\log g$ & 4.39\,$\pm$\,0.03 & From \rs\ and $M_{\rm{S}}$ \\ 
  Luminosity & 0.09\,$\pm$\,0.02 $L_{\odot}$ & \cite{plavchan2009} \\ 
  Age & 22\,$\pm$\,3\,Myr & \cite{mamajek2014} \\
  \pr & 4.86\,$\pm$\,0.01\,d &  \citetalias{plavchan2020} \\   
  $i_{\rm{orb}}$ & 89.5\,$\pm$\,0.4\degr & Planet orbit inclination of \citetalias{plavchan2020} \\
  \vs & 7.8\,$\pm$\,0.3\,\kms & From \rs, \pr\ and $i_{\rm{rot}}$\,=\,$i_{\rm{orb}}$ \\
    \hline
    \end{tabular}
    \label{tab:star_prop}
\end{table}

\section{Radial velocity analysis}\label{sec:rv_an}

\subsection{RV measurement}\label{sec:sec3_1}

\begin{figure}
    \centering
    \includegraphics[width=\linewidth]{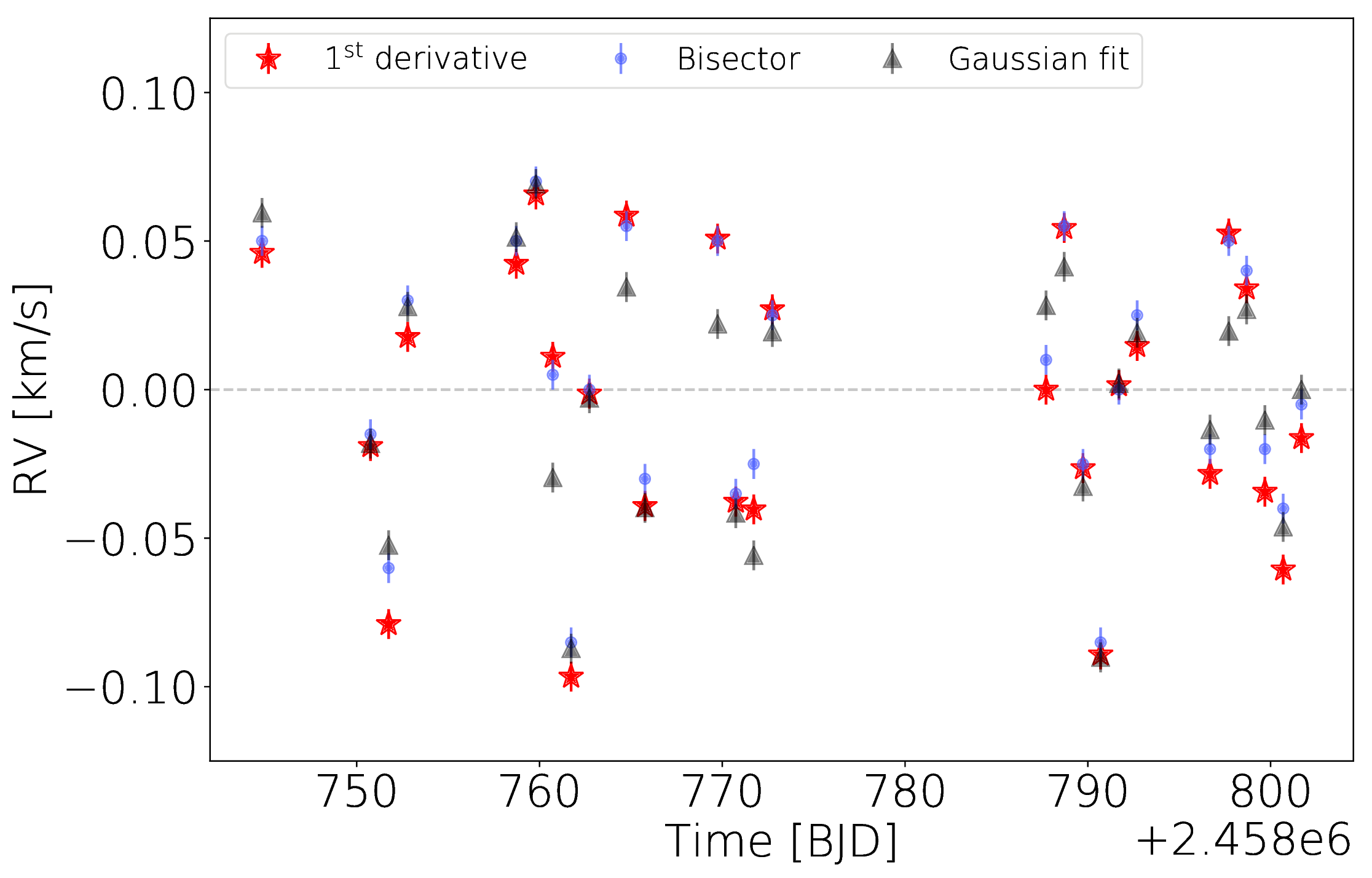}
    \caption{Median-subtracted RV time-series measured by (i)~fitting a Gaussian function to the Stokes $I$ LSD profiles (black triangles), (ii)~computing the median value of the bisector for each observation (blue dots), and (iii)~linearly adjusting the first derivative of a Gaussian function to the median-subtracted Stokes $I$ LSD profiles (red stars). All RV time-series exhibit similar dispersion of 45\,\ms\ RMS.}
    \label{fig:rvs}
\end{figure}


We carry out RV measurements of \aum\ by jointly fitting a Gaussian function on top of a linear continuum (4 parameters altogether) to each Stokes $I$ LSD profile extracted with the empirical mask and truncated at $\pm$31\,\kms\ from the line center, located at $\sim$4.45\,\kms. The slanted continuum used in the model accounts for residual slopes in the spectrum that could bias the RV measurement process if ignored. We find that the observed profiles are relatively well fitted by a Gaussian function with a median full-width at half maximum (FWHM) of 15.8\,\kms\ (see the RV measurements in Tab.~\ref{tab:list_obs}). We consider the resulting RV time-series to be the reference one for the following RV analysis. We implemented two additional methods to measure RVs from the Stokes $I$ LSD profiles. For the first one, we calculate the bisector of each line profile, corrected from residual slopes in the continuum beforehand, using the method introduced in \citet{gray1984} and \citet{queloz2001}. RVs are derived by computing the median bisector between 20 and 95\% of the full line depth (counting from the continuum). We also compute the velocity span, $V_{\rm{S}}$, known to be a reliable proxy of stellar activity RV signals \citep[e.g.,][]{queloz2001}, from average velocities on the top and bottom parts of the bisector (i.e., within 20-40 and 65-95\% of the line depth counting from the continuum, respectively; see Sec.~\ref{sec:act_ind} for the analysis of the $V_{\rm{S}}$ time-series). For the second method, we scale and subtract the median Stokes $I$ line, $\bar{I}$, from each LSD profile and linearly adjust the residuals with the first derivative of a Gaussian function fitted to $\bar{I}$ beforehand. The resulting RV time series are shown in Fig.~\ref{fig:rvs} and listed in Tab.~\ref{tab:App_A}. They all exhibit activity-induced fluctuations of $\sim$120\,\ms\ peak-to-peak and dispersions of $\sim$45\,\ms\ RMS, $\sim$2.5$\times$ lower than typical dispersions obtained with HARPS/HIRES spectrographs in \citetalias{plavchan2020} and similar to the typical dispersion observed in nIR RV observations of \aum\ \citep[e.g.,][and \citetalias{plavchan2020}, using respectively CSHELL and iSHELL spectrographs]{gagne2016}. The RV time-series exhibit median differences of $\sim$9\,\ms\ RMS between each other.



\subsection{Modeling the RV time-series}\label{sec:sec3_2}

We model each median-subtracted RV times-series, \vr, as the sum of a planetary RV signature, \vp, a stellar activity RV signal, \vj, and a constant offset, \voo, such that

\begin{eqnarray}
V_{\rm{r}} (t) = V_{\rm{p}} (t) + V_{\rm{j}} (t) + V_{0} + \epsilon (t),
\label{eq:rv_mod}
\end{eqnarray}

\noindent
where $\epsilon (t)$\,$\sim$\,$\mathcal{N}(0,\sigma(t))$, $\sigma(t)$ being the 1$\sigma$-uncertainty on the data point at time $t$, equal to 5\,\ms\ in our data set. Assuming a circular orbit for \aum~b \citepalias[based on the transit curve analysis of][]{plavchan2020}, the RV signal induced by the planet is simply described by

\begin{eqnarray}
V_{\rm{p}} (t) = -K_{\rm{p}} \sin \left[ 2 \pi\:  \frac{t - T_{0}}{P_{\rm{orb}}} - 2 \pi\: \phi_{\rm{p}} \right],
\label{eq:rv_p}
\end{eqnarray}

\noindent
where \kp\ is the semi-amplitude of the planetary signal, \porb, the orbital period of the planet and, \php, the orbital phase correction with respect to \tze/\porb\ (\tze\ being our reference time, set to the 2019 July 17 transit time, see Sec.\ref{sec:data_red}). We assume that the planet orbital period is known from photometry \citepalias[\porb\,=\,8.46321\,$\pm$\,0.00004\,d; see][]{plavchan2020}, and that the planetary signal is perfectly phased from the mid-transit time (i.e., \php\,=\,0), leaving \kp\ as the only free parameter of the planetary model. We use Gaussian-Process Regression \citep[GPR][]{rasmussen2006}, to model the stellar activity RV signal. Similarly to the RV analysis of \citetalias{plavchan2020}, we assume a quasi-periodic covariance kernel, $k$, relying on a vector of 4 hyperparameters, \hypv, and already known to accurately describe activity-induced fluctuations in RV curves for mature stars \citep{haywood2014,rajpaul2015}, as well as for PMS stars \citep[e.g.,][]{donati2016,yu2017,klein2020}, and given by

\begin{eqnarray}
\mathrm{k}(t_{i},t_{j}) = \mathrm{\theta}_{1}^{2} \exp \left[ - \frac{(t_{i}-t_{j})^{2}}{\mathrm{\theta}_{2}^{2}} - \frac{\sin^{2} \frac{\pi (t_{i}-t_{j})}{\mathrm{\theta}_{3}}}{\mathrm{\theta}_{4}^{2}} \right]
\label{eq:cov_fct}
\end{eqnarray}

\noindent
where $t_{i}$ is the time associated to observation $i$, and where $\theta_{1}$ to $\theta_{4}$ are respectively the amplitude, decay time, recurrence time-scale and smoothing factor of the Gaussian Process (GP). To investigate whether our model is self-consistent regarding noise statistics, we proceed as in \citet[][]{suarez2020} and introduce, as an additional free model parameter, a term describing the excess of uncorrelated noise, $S$, accounting for potential systematics in the data beyond the assumed noise level of 5\,\ms\ (e.g., due to stellar variability or residuals of telluric correction). The free parameters are jointly estimated by maximizing the posterior density of the model $p($\hypv,\kp,\voo,$S$|\vr$)$, sampled using the \textsc{emcee} Markov Chain Monte Carlo (MCMC) sampler \citep{Foreman-Mackey2013}, in the Bayesian framework. We run our MCMC on 5000 iterations of 100 walkers, i.e., significantly larger than the typical autocorrelation time of the chain, found to be $\sim$100 iterations in our case. The prior densities adopted for the estimation process are given in Table~\ref{tab:priors}. For \kp, $\theta_{1}$ and $S$, we use a modified Jeffreys prior \citep{gregory2007}, whose knee is set at the mean 1$\sigma$-uncertainty on our RV data, $\bar{\sigma}$\,=5\,\ms, so that the prior densities are uniform (resp. log-uniform) when the parameters are small (resp. large) compared to $\bar{\sigma}$ \citep[see][]{haywood2014}. Given the low number of data points in our RV time-series, we fix the values of $\theta_{2}$ and $\theta_{4}$ to 100\,d and 0.4 respectively
\citepalias[i.e., close to those quoted in][]{plavchan2020}, found to be consistent with those derived from the longitudinal field analysis\footnote{Fixing $\theta_{2}$ and $\theta_{4}$ to values differing by $\sim$1$\sigma$ than those reported in \citetalias{plavchan2020} only marginally impacts the recovered parameters of the model.} (see Sec.~\ref{sec:act_ind}). The median and 1$\sigma$ error bars on the model parameters are computed from the posterior density after removing the first 200\,000 iterations (i.e., much larger than the autocorrelation time of the chain).

\begin{table}
\centering
\caption{\label{tab:priors}Prior densities used in the Bayesian MCMC sampling of the posterior density of our RV model. Parameters written in bold are locked in the MCMC process except when otherwise indicated in the text. The knee of the modified Jeffreys prior is chosen to be the mean 1$\sigma$-uncertainty on the RV measurements (in our case, $\bar{\sigma}$\,=\,5\,\ms). }
\begin{tabular}{ccc}
\hline
Quantity & Parameter & Prior \\
\hline
Semi-amplitude of the planet signal & \kp &  Modified Jeffreys ($\bar{\sigma}$) \\
Planet orbital period &  \porb &\textbf{8.46321\,d} \\
Planet orbital phase & \php & \textbf{0.0} \\
GP amplitude &  $\theta_{1}$  & Modified Jeffreys ($\bar{\sigma}$) \\
GP decay time & $\theta_{2}$  & \textbf{100\,d} \\
GP recurrence period & $\theta_{3}$  &  $\mathcal{U}(4.5,5.3)$ [d] \\
GP smoothing parameter & $\theta_{4}$ &  \textbf{0.4} \\
Constant velocity offset & \voo &  $\mathcal{U}(-100,100)$ [\ms] \\
Excess of uncorrelated noise & $S$ & Modified Jeffreys ($\bar{\sigma}$) \\
\hline
\end{tabular}
\end{table}

The significance of the planet detection in the RV time-series is estimated through a Bayesian comparison of models $M_{1}$ and $M_{0}$, searching for 1 and 0 planet in the RV time-series, respectively. This is done by computing the so-called posterior odds ratio $p\left( M_{1}|V_{\rm{r}}\right)/p\left( M_{0}|V_{\rm{r}}\right)$, that is proportional to the Bayes factor (BF), i.e., the ratio of marginal likelihoods of models $M_{1}$ and $M_{0}$ \citep[see Sec.~3 in][]{diaz2014}. BF is estimated by applying the method introduced in \citet{Chib2001} as described in \citet{haywood2014} and \citet{klein2019}. Following \citet{Jeffreys1961}, a BF larger than 150 (i.e., 5 in $\log$, where $\log$ refers to the natural logarithm) would theoretically be interpreted as a fair detection of the planet in the RV time-series, while, in practice $\ln$\,BF beyond 10 seems to be a more reliable threshold to claim a definite detection \citep[see the recommandations of][on the interpretation of BF values]{nelson2018}.

\subsection{Results}\label{sec:sec3.3}

\begin{figure}
    \centering
    \includegraphics[width=\linewidth]{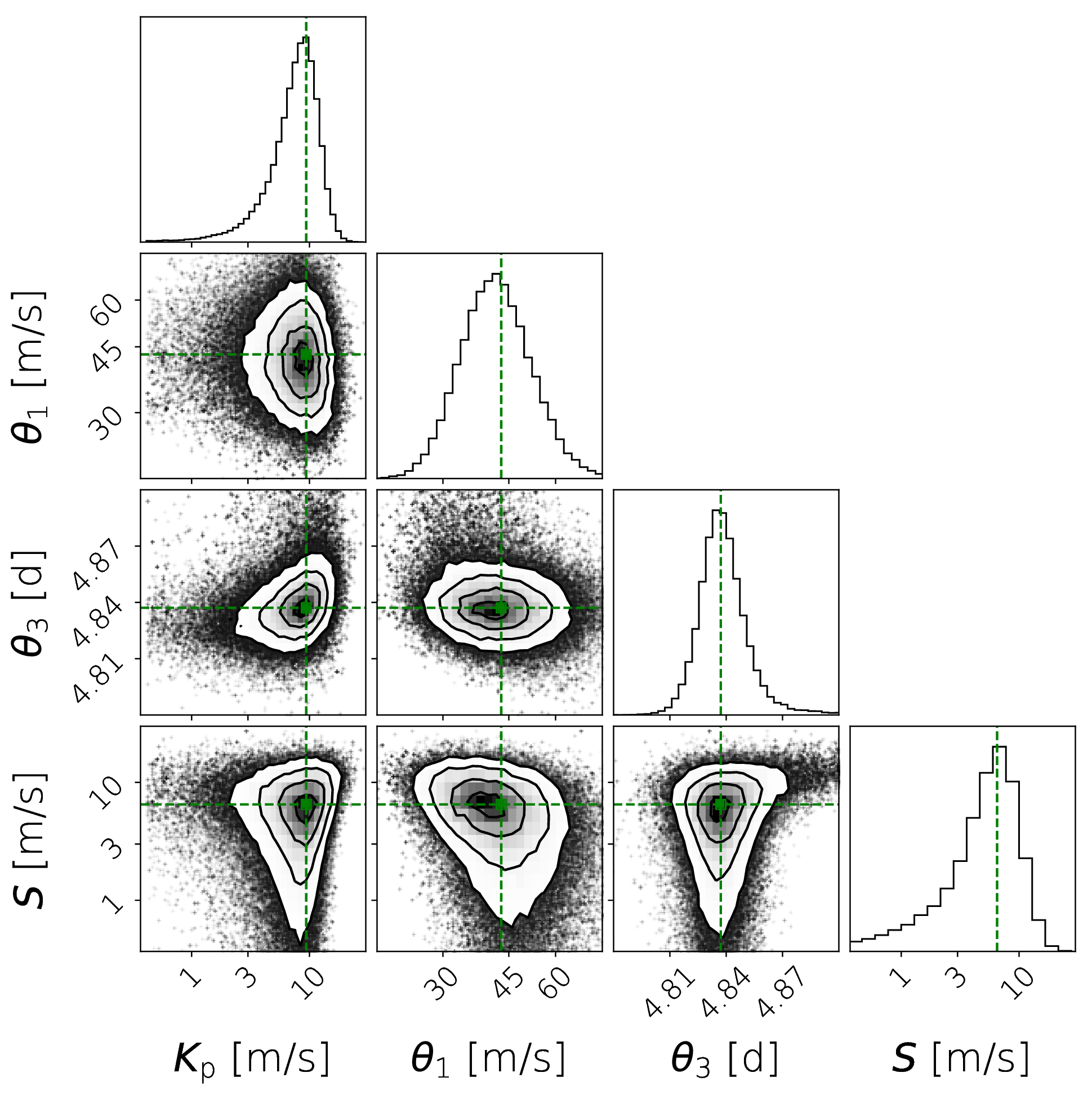}
    \caption{Posterior density resulting from the MCMC sampling of the planet and stellar activity parameters using the method described in Sec.~\ref{sec:sec3_2}. The concentric circles within each panel indicate the 1, 2 and 3$\sigma$ contours of the distribution. The green dashed lines indicate the parameters that maximize the posterior density. This figure was made using the \textsc{corner} python module \citep{Foreman-Mackey2016}. Note that the distributions of \kp, $\theta_{1}$ and $S$ are shown as $\log$-plots.}
    \label{fig:corner_rv}
\end{figure}

{\renewcommand{\arraystretch}{1.25} 
\begin{table*}
    \centering
    \caption{Results of the fit to the RV time-series in case of RVs measured by (i)~fitting a Gaussian function to the raw Stokes $I$ LSD profiles (lines~1 to~5), (ii)~computing the median value of the bisector (line~6), and (iii)~fitting the first derivative of a Gaussian function to the median-subtracted Stokes $I$ LSD profiles (line~7). For comparison purposes, we also list the results of the modeling of the RV time-series obtained by fitting a Gaussian function to the rescaled Stokes $I$ LSD profiles prior to the brightness reconstruction in lines~8 and~9 (namely $\boldsymbol{I_{\rm{f}}}$), and of the ZDI reconstruction in line~10 (see Sec.~\ref{sec:bright_im}). In column 2, we give the typical RV uncertainties $\bar{\sigma}$, set to 5\,\ms\ in all cases except for line 5 where we use instead the estimated photon noise $\sigma_{\rm{ph}}$. Columns~3 and~4 list the hyperparameters that maximize the posterior density (with $\pm$1$\sigma$ uncertainties; in all cases, $\theta_{2}$ and $\theta_{4}$ are respectively fixed 100\,d and 0.4.). Columns~5 to~8 give the best estimates of \kp, \php, \porb\ and \mpp, computed using the stellar parameters given in Tab.~\ref{tab:star_prop} and \kp. Columns~9 and~10 give the best estimates of the constant velocity offset \voo\ and the excess of uncorrelated noise in our data set, $S$. All parameters are written in bold when non optimized in the MCMC process. Finally, columns~11 and 12 indicate the RMS of the residuals of the fit and the BF in favour of \aum~b in our RV time-series.}
    \label{tab:results_rv}

    \begin{tabular}{cccccccccccc}
    \hline
    Case &  $\bar{\sigma}$  &  $\theta_{1}$ & $\theta_{3}$  & \kp  & \php & \porb &  \mpp & $V_{0}$  & $S$ & RMS & $\ln$ BF \\
    & [\ms] & [\ms] &  [d] & [\ms] & & [d] & [\mpear] & [\ms] & [\ms]  & [\ms]  & \\
    \hline
      Gaussian fit (reference)  & 5.0  &  47$^{+11}_{-8}$  & 4.836 $\pm$ 0.009  & 8.5$^{+2.3}_{-2.2}$ & \textbf{0.0} & \textbf{8.46321}  & 17.1$^{+4.7}_{-4.5}$ & 2$^{+20}_{-19}$ & \textbf{0.0} &  3.0 & 5.6 \\

      Gaussian fit (\php\ free)  & 5.0 &  45$^{+11}_{-8}$ &  4.836 $\pm$ 0.009 &  8.2$^{+2.4}_{-2.3}$ & -0.03 $\pm$ 0.04 & \textbf{8.46321} & 16.5$^{+4.9}_{-4.7}$ & 3$^{+20}_{-19}$ & \textbf{0.0} & 3.1 & 6.3 \\ 
      Gaussian fit (\porb\ free)  & 5.0 & 46$^{+10}_{-8}$ & 4.834 $\pm$ 0.009 & 8.5 $\pm$ 2.3 & \textbf{0.0} & 8.5 $\pm$ 0.03 d & 17.1 $\pm$ 4.7 & -1 $\pm$ 19 & \textbf{0.0} & 3.7 & 7.3 \\
      
      Gaussian fit ($S$ free)   & 5.0  & 43$^{+11}_{-8}$  & 4.84 $\pm$ 0.01   & 9.3$^{+3.2}_{-3.0}$ & \textbf{0.0} & \textbf{8.46321}  & 18.7$^{+6.5}_{-6.1}$ &  0$^{+19}_{-18}$    &  6.0$^{+3.8}_{-3.1}$  & 4.4 & 4.5 \\
      
      Gaussian fit ($S$ free)   & $\sigma_{\rm{ph}}$  & 40$^{+11}_{-7}$  & 4.84 $\pm$ 0.01   & 9.3$^{+3.5}_{-3.2}$ & \textbf{0.0} & \textbf{8.46321}  & 18.7$^{+7.1}_{-6.5}$ &  2$^{+20}_{-19}$   &  7.2$^{+3.2}_{-2.0}$  & 4.3 & 4.1 \\    
      
      \hline
      Bisector  & 5.0  & 80$^{+20}_{-15}$ & 4.85 $\pm$ 0.01 & 8.6$^{+2.5}_{-2.4}$ & \textbf{0.0} & \textbf{8.46321} & 17.3$^{+5.1}_{-4.9}$ & -13 $\pm$ 36 &  \textbf{0.0} & 3.3 & 5.7 \\
      1$^{\rm{st}}$ derivative  & 5.0 & 91$^{+21}_{-17}$ & 4.85 $\pm$ 0.01 & 7.8$^{+2.6}_{-2.5}$ & \textbf{0.0} & \textbf{8.46321} & 15.7$^{+5.3}_{-5.1}$ &  -17 $\pm$ 40  &  \textbf{0.0}  & 3.8 & 4.0 \\
     \hline
     Gaussian fit to $\boldsymbol{I_{\rm{f}}}$  & 5.0 & 45$^{+11}_{-8}$ & 4.85 $\pm$ 0.01 & 10.1 $\pm$ 2.3 & \textbf{0.0} & \textbf{8.46321} &  20.3 $\pm$ 4.7 & -15$^{+19}_{-18}$  & \textbf{0.0}  & 3.0 & 7.3 \\
     Gaussian fit to $\boldsymbol{I_{\rm{f}}}$ ($S$ free)  & 5.0 & 38$^{+11}_{-7}$ & 4.86 $\pm$ 0.02 & 10.1$^{+3.1}_{-2.9}$ & \textbf{0.0} & \textbf{8.46321} & 20.3$^{+6.4}_{-6.3}$ & -15$^{+19}_{-18}$ &  6.4$^{+3.2}_{-2.9}$ & 5.0 & 4.8 \\
     ZDI reconstruction  & 5.0 & -- & -- & 9.7 $\pm$ 2.5 & \textbf{0.0} & \textbf{8.46321} & 19.1 $\pm$ 5.1 & -- & -- & 8.7 & -- \\
    \hline
    \end{tabular}
\end{table*}}

\begin{figure}
    \centering
    \includegraphics[width=\linewidth,height=8cm]{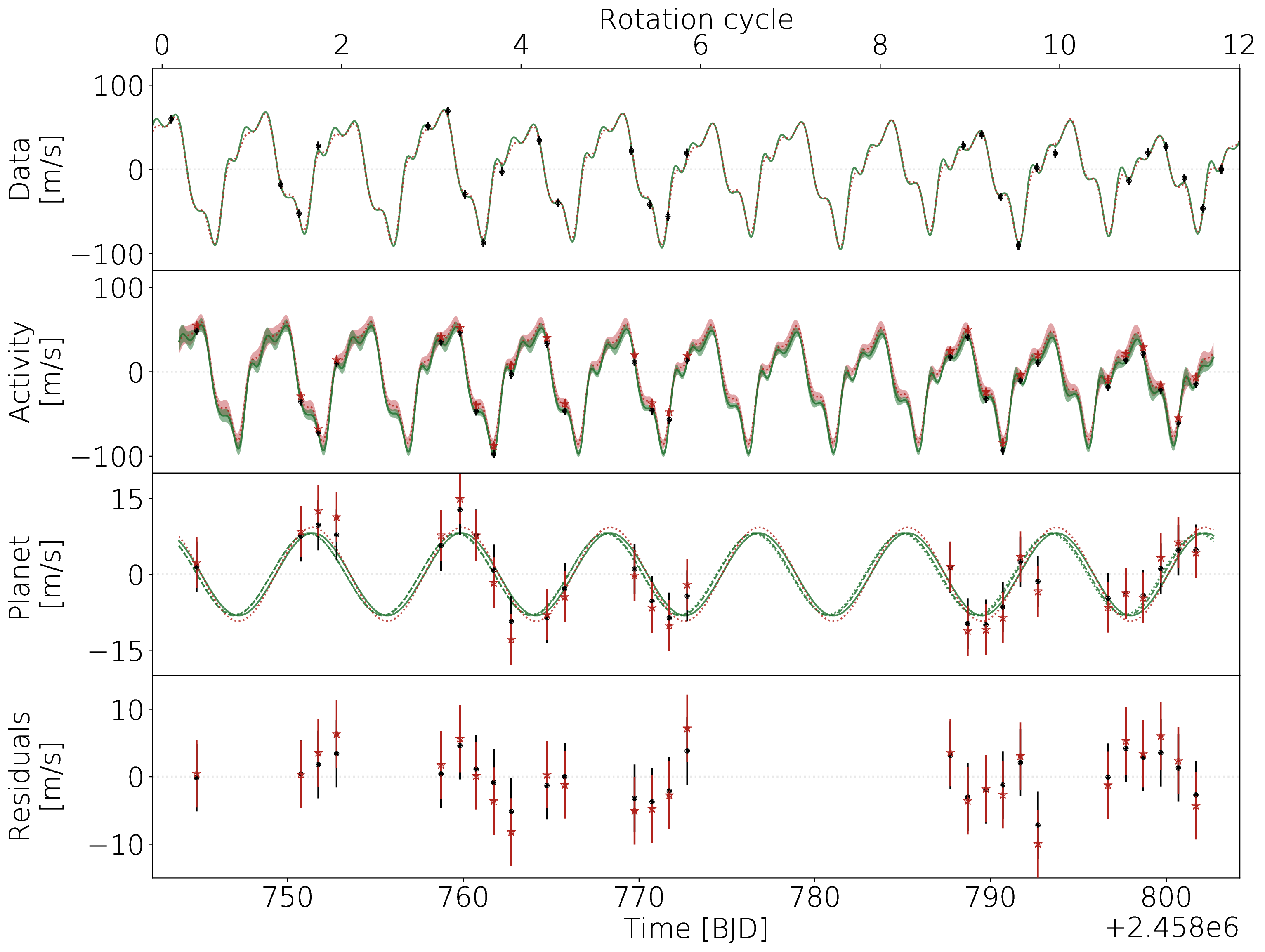}
    \caption{Best fit to the reference RV time-series. From top to bottom, we show the raw RV time-series, the stellar activity contribution, the planetary RV signature and the residuals after subtracting stellar and planetary predictions from the data. In each panel, the green solid line and red dotted line (resp. the $\pm$1$\sigma$ error bands of the GP prediction in panel 2) show the best prediction of the model respectively without and with the additional uncorrelated noise $S$ fitted by the MCMC process. The data points (red stars and black dots when $S$ is respectively optimized in the estimation process and fixed to 0\,\ms) in panels 2 and 3 and obtained by respectively subtracting the reconstructed planet and stellar activity RV signals from the raw RVs. In panel 3, the dashed and dotted green lines indicate the planet RV signal respectively obtained when \php\ and \porb\ are regarded as free parameters in the MCMC process. The residuals of the fit, shown in panel 4, exhibit respective RMSs of 3.0 or 4.4\,\ms\ when $S$ is assumed to be null or fitted in the MCMC process.}
    \label{fig:pred_rv}
\end{figure}

\begin{figure}
    \centering
    \includegraphics[width=\linewidth]{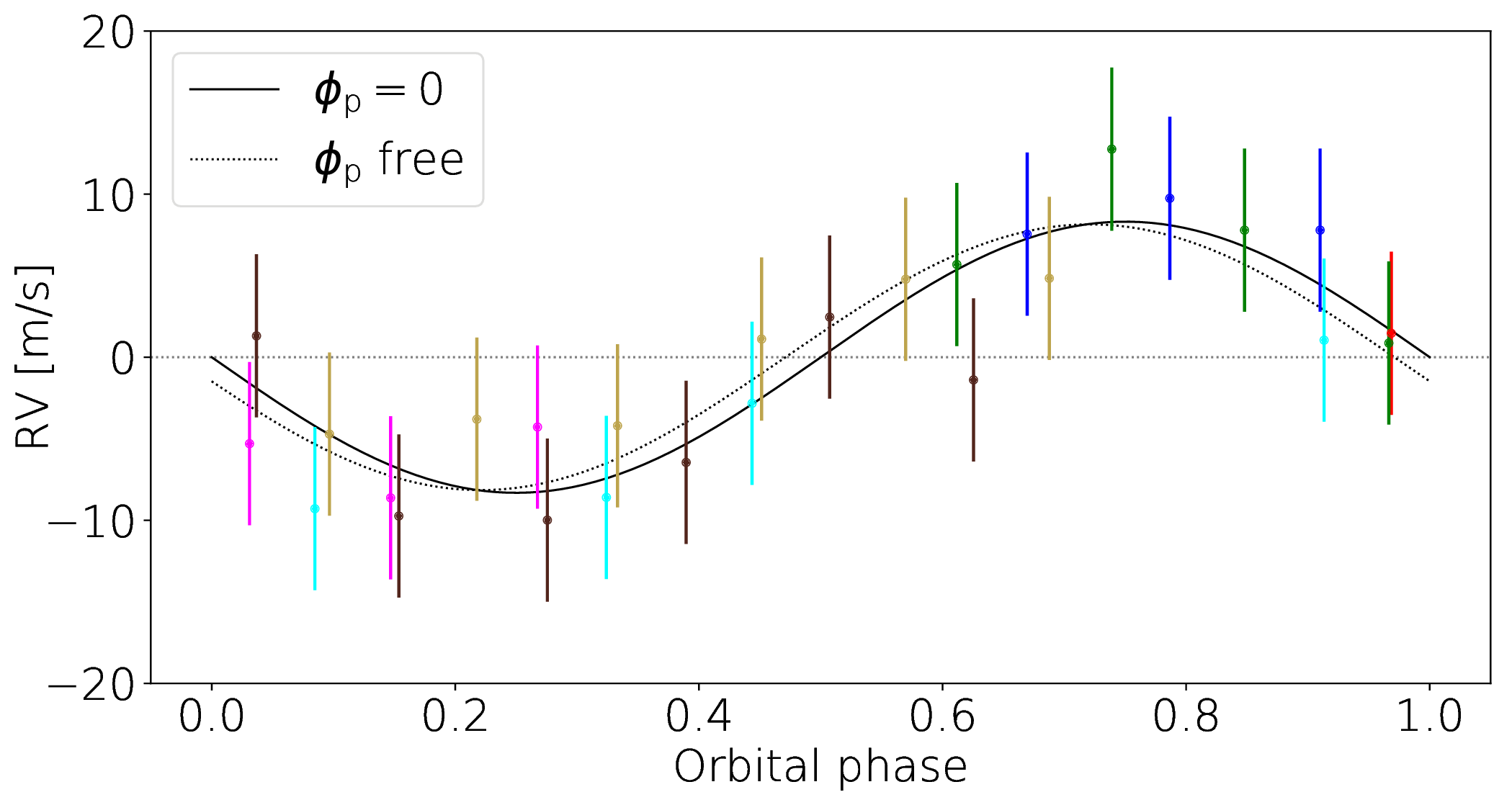}
    \caption{Activity-subtracted RV time-series (dots) folded to \porb\ using \tze\ as reference time. The subtracted activity RV signal is estimated assuming \php\,=\,0 and $S$\,=\,0\,\ms. Data points of same color belong to the same orbital phase. We plot in solid and dotted black lines, the planet RV signatures obtained assuming \php\,=\,0 and optimizing \php\ in the MCMC process, respectively.}
    \label{fig:phase_fold_rv}
\end{figure}

We fit the model described in Sec.~\ref{sec:sec3_2} to the different RV time-series obtained in Sec.~\ref{sec:sec3_1}, assuming for a start that $S$\,=\,0\,\ms. The main parameters of the estimation processes are given in Tab.~\ref{tab:results_rv}. The MCMC process is efficiently converging in all cases, leading to Gaussian-like posterior densities for all parameters, as illustrated in Fig.~\ref{fig:corner_rv}. The best fit to our reference RV time-series is shown in Fig.~\ref{fig:pred_rv}. In this case, we report a 3.9$\sigma$ detection of a planetary signal of \kp\,=8.5$^{+2.3}_{-2.2}$\,\ms\ at the orbital period reported in \citetalias{plavchan2020}. Consistent results are also obtained when RVs are measured using the two additional methods presented in Sec.~\ref{sec:sec3_1} (see the results in Tab.~\ref{tab:results_rv}). With BF values lying close or above the fair detection threshold (of $\ln$\,BF\,=\,5) in most cases, in particular in the reference one (first line of Tab.~\ref{tab:results_rv}), we can in principle conclude that the planet RV signal is formally detected, although we caution that more data are needed to raise $\ln$\,BF beyond 10 for a definite detection \citep{nelson2018}. Leaving \php\ or \porb\ as free parameters of the MCMC process yields similar results (see Tab.~\ref{tab:results_rv}), further confirming that the recovered planetary signal is well-phased with the photometric transits reported in \citetalias{plavchan2020} (see also the orbit-folded time-series shown in Fig.~\ref{fig:phase_fold_rv}). As expected 
from the fact that our GP is applied to the median-subtracted RV time-series, \voo\ is poorly-constrained by our data set, and only marginally impacts the recovered planet and stellar activity signals.



When $S$ is now left as a free parameter of the model, we find $S$\,=\,6.0$^{+3.8}_{-3.1}$\,\ms\ in our reference RV time-series (see line~4 of Tab.~\ref{tab:results_rv}), suggesting the presence of excess uncorrelated noise in our data\footnote{A consistent value of $S$\,=\,7.2$^{+3.2}_{-2.0}$\,\ms\ is obtained when using the photon noise alone as the formal RV uncertainty rather than the conservative 5\,\ms\ assumed in this study (see line 5 of Tab.~\ref{tab:results_rv})}. However, by applying our MCMC process to synthetic RV time-series mimicking our data, we find that the low number of points in our RV time-series makes it impossible for the estimator to fully disentangle the respective contributions of the GP and of the excess uncorrelated noise, with the GP signal being partly transferred to $S$ as a result (more information on these simulations are provided in Appendix~\ref{app:B}). As a consequence, $S$ is likely over-estimated when inferred from our limited data set.


Modeling the RV time-series obtained by computing the median value of the bisector and by fitting the first derivative of a Gaussian function to the median-subtracted Stokes $I$ LSD profiles yields consistent estimates of the planet mass (see lines 6 and 7 of Tab.~\ref{tab:results_rv}). However, the excess uncorrelated noise $S$ takes values as high as 15\,\ms\ in both cases, suggesting that, in the case of active stars, both alternative methods are noisier than our reference RVs and expected to yield less accurate/reliable estimates of the planet mass.

\begin{figure}
    \centering
    \includegraphics[width=\linewidth]{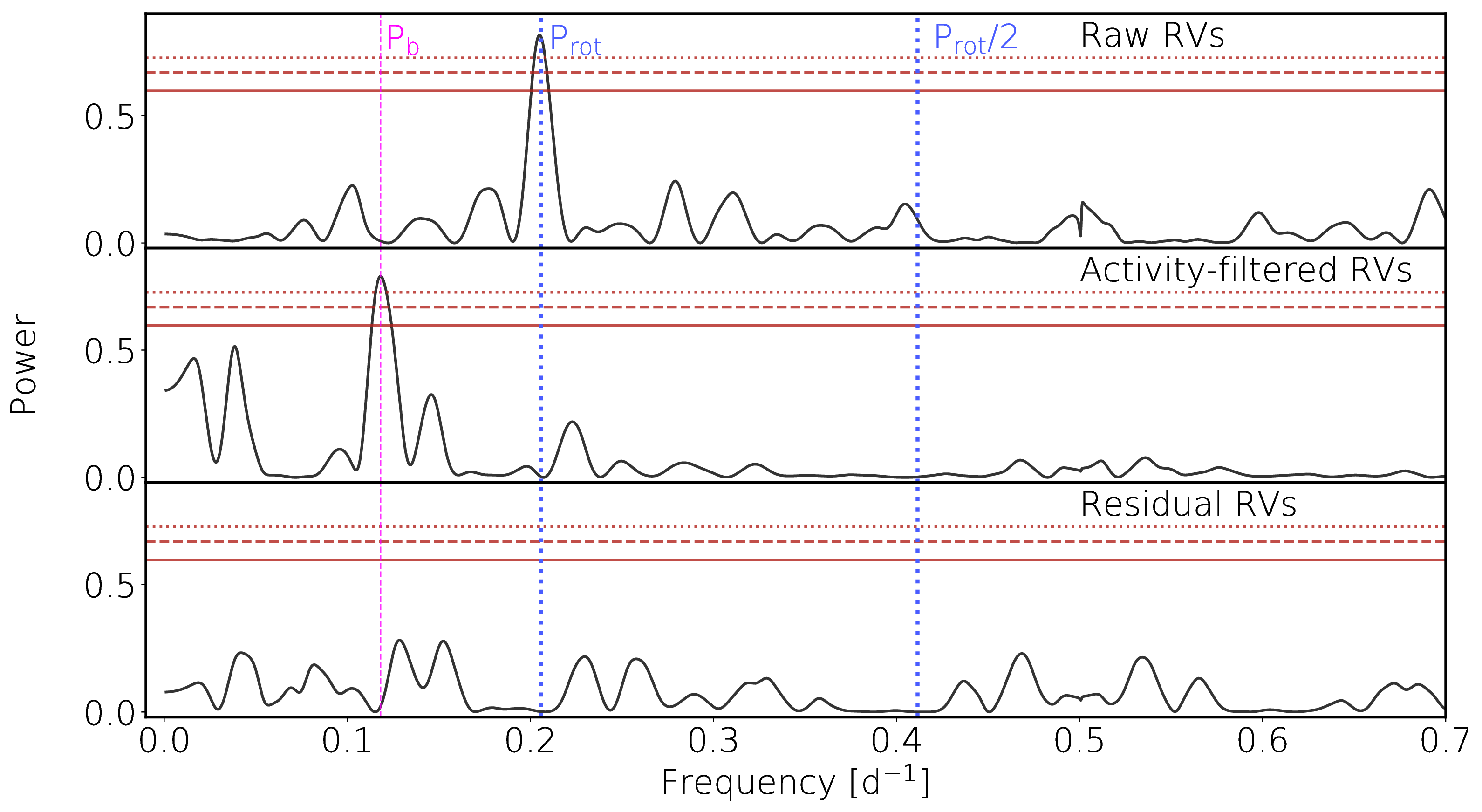}
    \caption{GLS periodograms of our reference RV time-series (top panel), the RVs obtained after subtracting the GP prediction of the stellar activity signal (\php\,=\,0, $S$\,=\,0\,\ms; middle panel), and the residual RVs (bottom panel). The horizontal solid, dashed and dotted red lines indicate false alarm probabilities of respectively 10, 1 and 0.1\%, computed using the method described in \citet{zechmeister} and the \textsc{PyAstronomy} python package \citep{pya}. In each panel, we indicate the frequencies corresponding to \aum~b (P$_{\rm{b}}$) and the stellar rotation period (and its first harmonic) in magenta dashed lines and blue dotted lines, respectively.}
    \label{fig:period_rv}
\end{figure}

To independently assess the reliability of the detected planet signal, we also computed a Generalized Lomb-Scargle periodogram \citep[GLS;][]{zechmeister} of the RV time-series obtained by subtracting the modeled stellar activity signal from the raw RV time-series (see Fig.~\ref{fig:period_rv}). We find a prominent peak at \porb\ that lies slightly above the 0.1\% false alarm probability level (FAP; see Fig.~\ref{fig:period_rv}). We do not find statistically-significant evidence for additional planets in our data set. The GP fit to the data features slopes in the modeled RV curve reaching up to $\sim$200\,\ms\,d$^{-1}$, with a median value of $\sim$0\,\ms\,d$^{-1}$ over the course of our observing run, and a standard deviation of 80\,\ms\,d$^{-1}$.


\begin{figure*}
    \centering
    \includegraphics[width=0.33\linewidth]{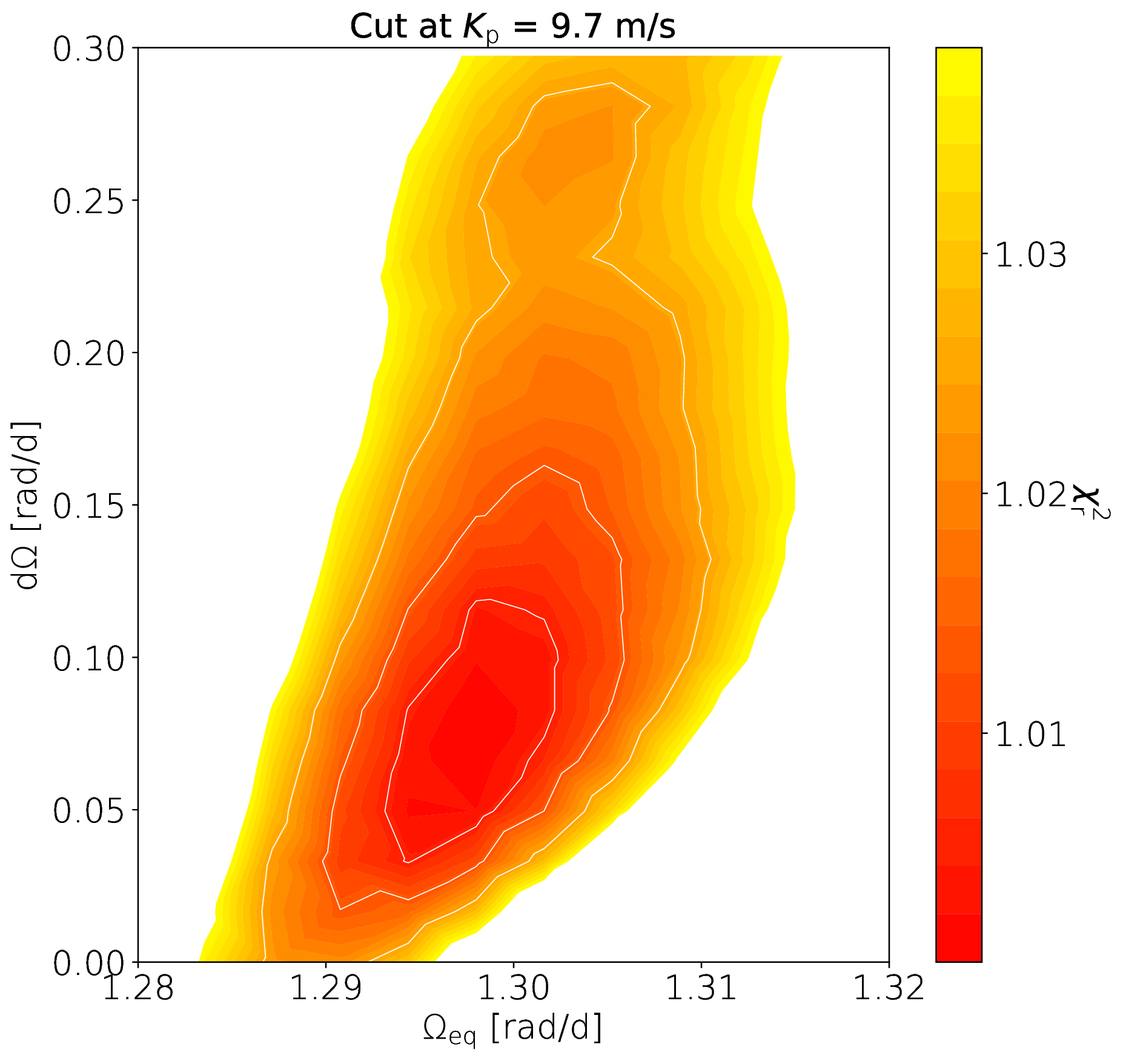}
    \includegraphics[width=0.33\linewidth]{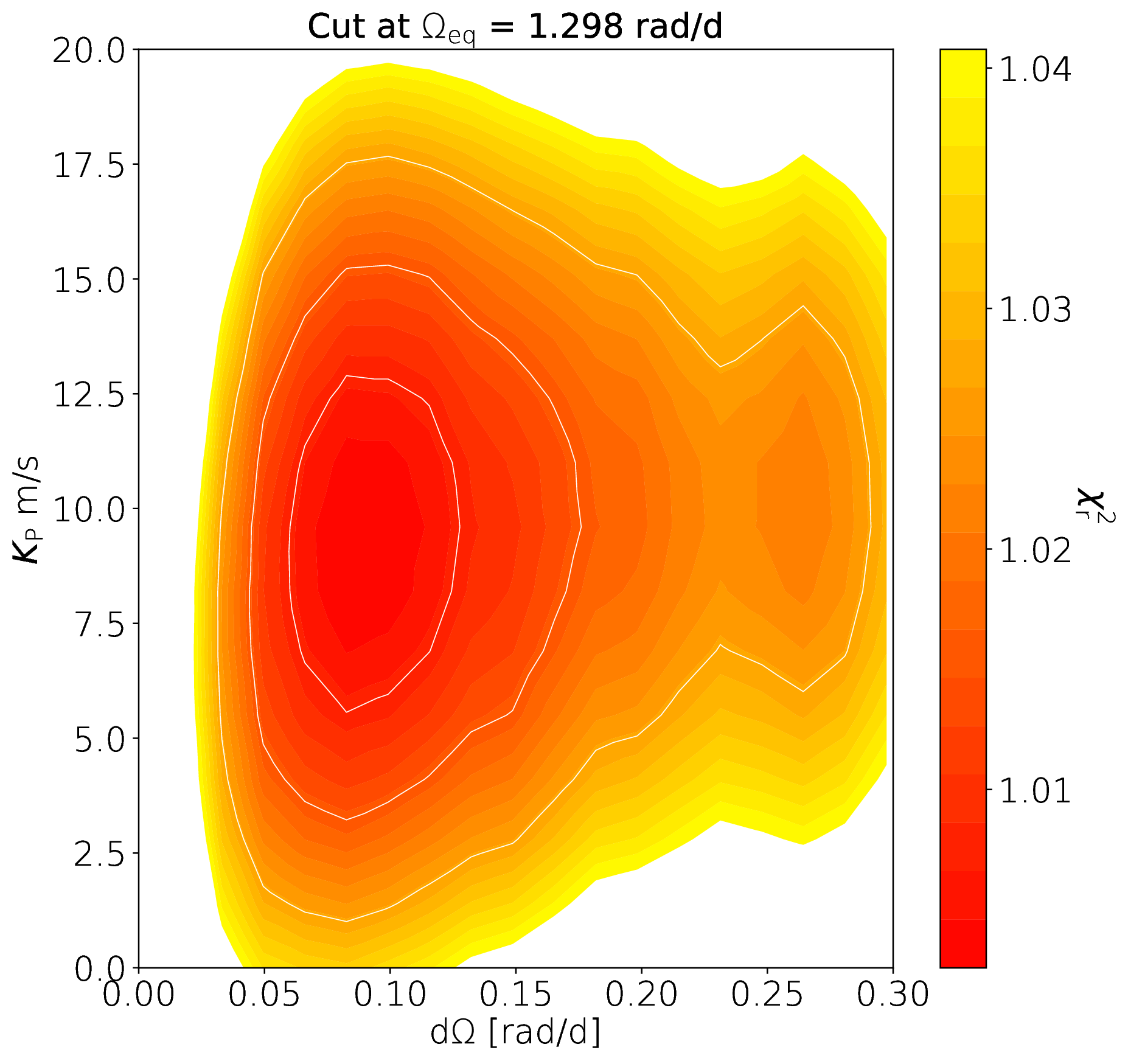}
    \includegraphics[width=0.33\linewidth]{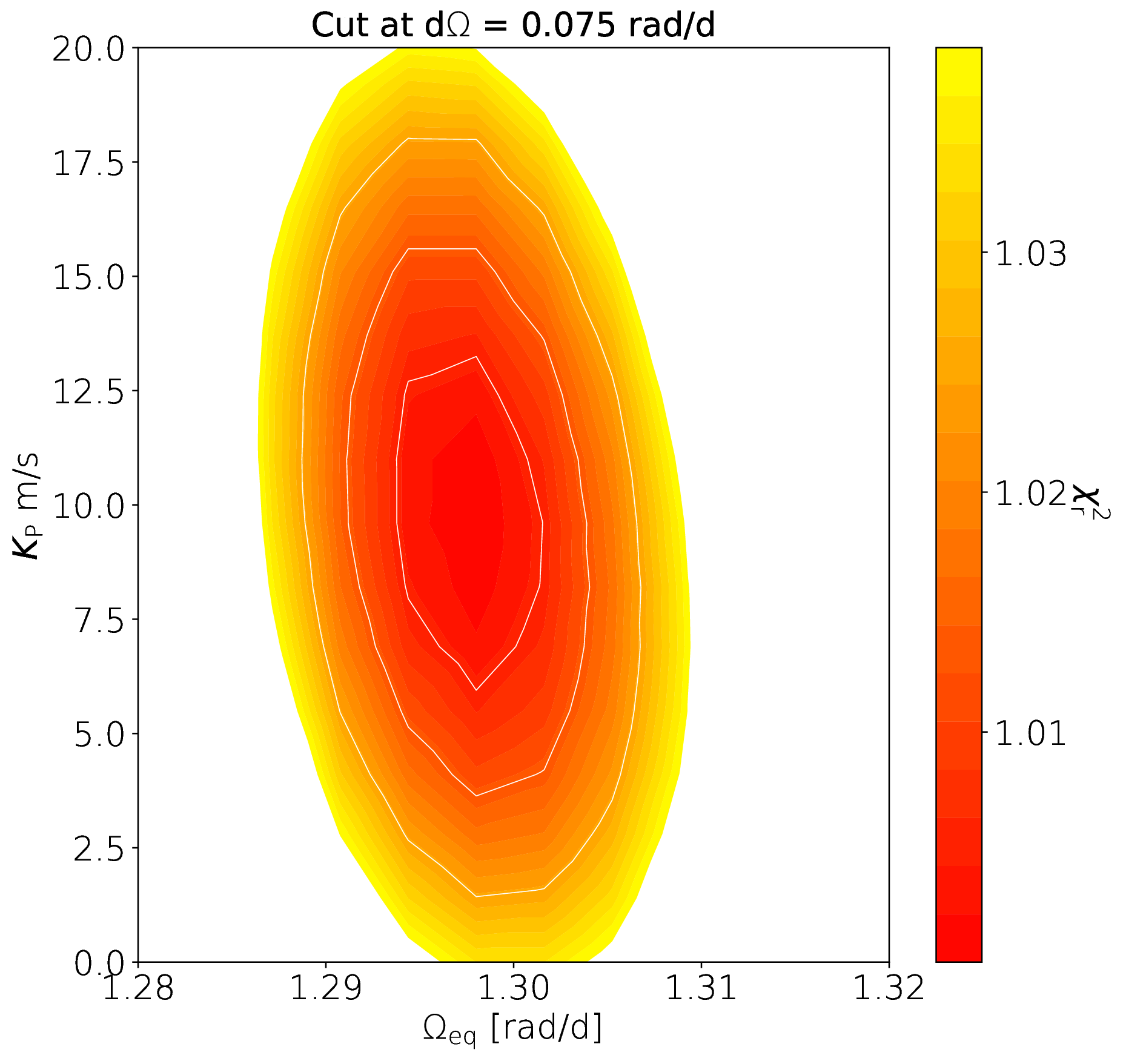}
    \caption{2D Cuts of the 3D \crr\ map obtained by performing brightness reconstructions of $\boldsymbol{I_{\rm{f}}}$ for a range of (\oeq, \dome, \kp). From right to left: variations of \crr\ in the (\oeq, \dome) space at \kp\,=\,9.7\,\ms, in the (\kp, \dome) space at \oeq\,=\,1.289\,rad/d and, in the (\oeq, \kp) space at \dome\,=\,0.075\,rad/d. In each panel, the 1, 2 and 3$\sigma$ levels are depicted by white solid lines.}
    \label{fig:3D_DR}
\end{figure*}

\section{Brightness imaging}\label{sec:bright_im}




\begin{figure}
    \centering
    \includegraphics[width=\linewidth]{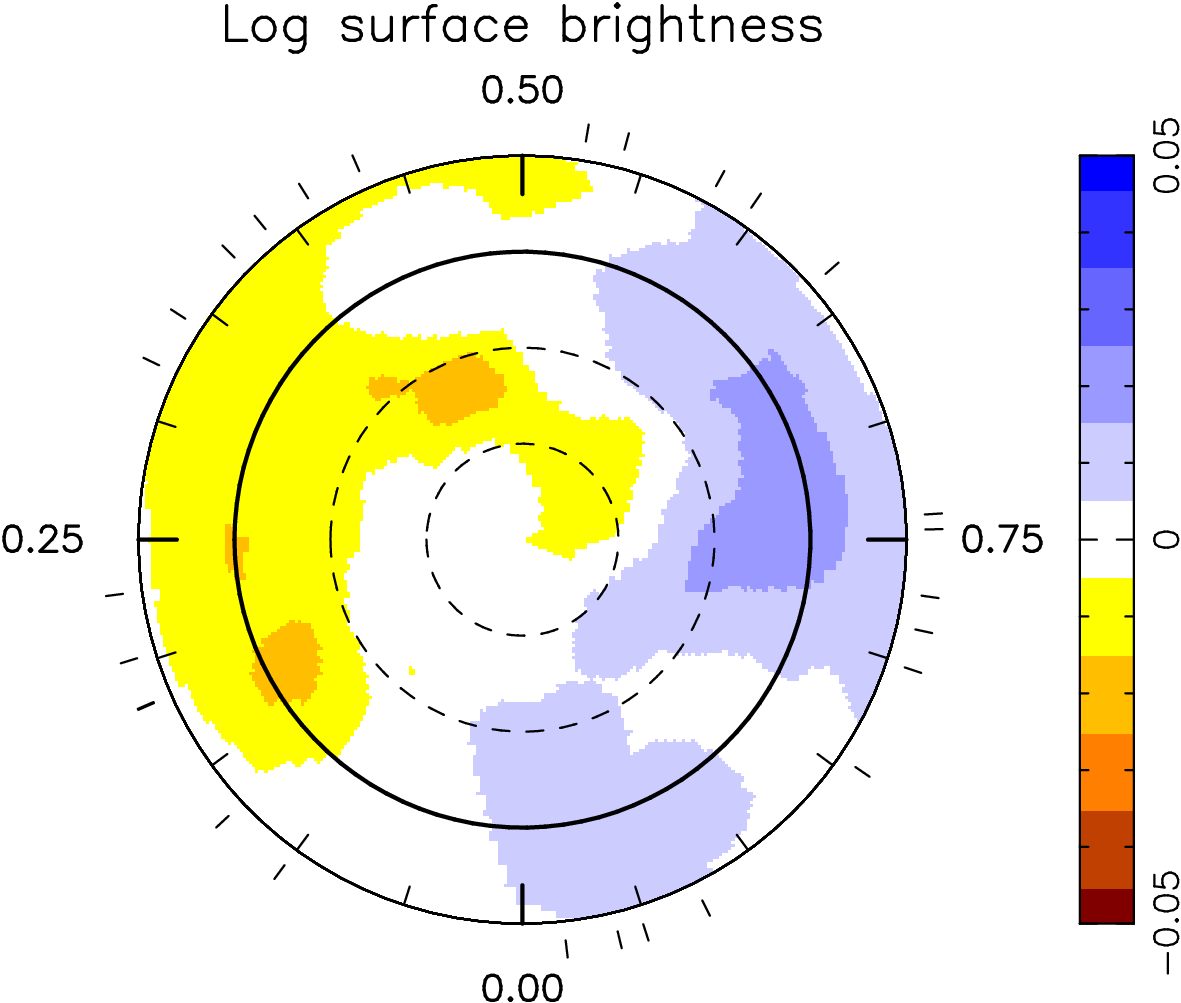}
    \caption{Flattened polar view of the surface brightness of \aum. The color scale depicts the logarithm of the surface brightness (relative to the quiet photosphere), with blue/brown regions standing for bright/dark features. The black circles indicate the stellar equator (bold line) and the -30\degr, 30\degr\ and 60\degr\ parallels (dashed lines). Ticks around the star mark the spectropolarimetric observations of the star.}
    \label{fig:surf_bright}
\end{figure}

\begin{figure}
    \centering
    \includegraphics[width=\linewidth]{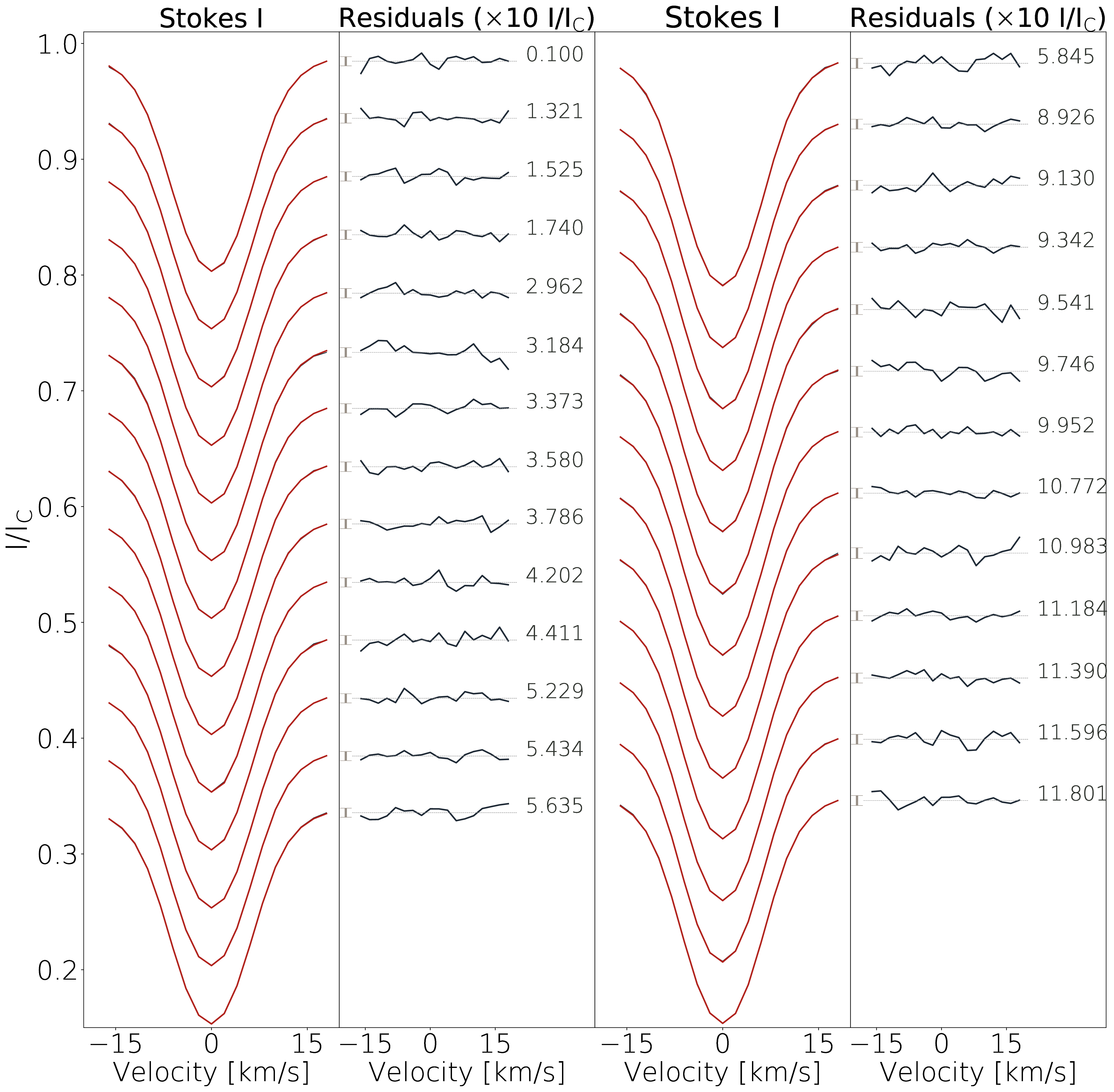}
    \caption{Panels 1 and 3: Best ZDI fit (red solid lines) to the observed Stokes $I$ LSD profiles (black solid lines) computed with the empirical line mask. Panels 2 and 4: Residual of fit in unit of $10 \times I/I_{\rm{C}}$. The $\pm$\,1$\sigma$ error bar (in unit of $10 \times I/I_{\rm{C}}$) and corresponding rotational cycle are indicated on both sides of each residual profile.}
    \label{fig:StokesI_Q}
\end{figure}

In this section, we propose to adapt the method introduced in \citet{petit2015} to simultaneously recover the barycentric motion due to the planet while inverting the Stokes $I$ LSD profiles with Zeeman Doppler imaging \citep[ZDI; e.g.,][]{semel1989,brown1991,donati1997b} into a surface brightness distribution of \aum. ZDI proceeds through an inversion process iteratively comparing the observed LSD profiles to simulated ones from a model star whose surface brightness distribution is updated until an adequate match to the observations is reached. More specifically, ZDI carries out a maximum entropy reconstruction of the surface brightness from the adjusted LSD profiles. ZDI samples the stellar surface with a dense grid of 5\,000 cells from which local Stokes $I$ profiles are computed using Unno–Rachkovsky’s analytical solution to the radiative transfer equations in a Milne–Eddington atmosphere. The local profiles are Doppler shifted and weighted by the relative brightness, limb darkening and stellar inclination, before being integrated over the visible stellar hemisphere. The time series of synthetic profiles are iteratively compared to the observed Stokes $I$ LSD profiles until reaching the maximum entropy solution at a given level of reduced $\chi^{2}$ (noted \crr).

To ensure that the code robustly and efficiently converges toward the maximum entropy solution, we start by correcting residual slopes with a linear fit to the extreme profile wings (as done when deriving RVs from a Gaussian fit to the LSD profiles, see Sec~\ref{sec:sec3_1}), so that observations optimally match modeled profiles in the continuum; we also homogenize (as usual in ZDI) the equivalent widths of all LSD profiles and truncate them at $\pm$18\,\kms\ from the line centre. Although \aum's axial inclination is likely close to 90\degr\ (see Tab.~\ref{tab:star_prop}), we impose a stellar inclination of $i_{\rm{zdi}}$\,=\,80\degr\ in the reconstruction, in order to prevent north-south degeneracy throughout the inversion process. From the stellar radius, rotation period and stellar inclination (assumed equal to the inclination of the planet orbit) listed in Tab.~\ref{tab:star_prop}, we find a projected rotational velocity \vs\,=\,7.8\,$\pm$\,0.3\,\kms, consistent with the width of the observed profiles and the value reported in \citet{weise2010}. Finally, we assume a linear limb-darkening law with a coefficient of 0.2 \citepalias[as in][]{plavchan2020}.


In stars like AU~Mic where the intrinsic profile at the surface of the star significantly contributes to the width of the observed LSD profile (as a result of the moderately low \vs\ of 7.8\,\kms), one has to ensure that the imaging code concentrates on modeling the profile variations rather than the profiles themselves \citep[e.g.,][]{hebrard2016}. In that aim, we first perform a maximum entropy reconstruction of the observed $\boldsymbol{I}$ profiles with ZDI, which yields an initial set of synthetic profiles $\boldsymbol{I_{\rm{s}}}$. We then correct the observed profiles by subtracting from $\boldsymbol{I}$ the median difference between $\boldsymbol{I}$ and $\boldsymbol{I_{\rm{s}}}$, and repeat the procedure until the median difference between $\boldsymbol{I}$ and $\boldsymbol{I_{\rm{s}}}$ is flat.

As a sanity check, we measure the RVs from our final Stokes $I$ LSD profiles (called $\boldsymbol{I_{\rm{f}}}$), found to differ from our reference RVs by no more than 4.5\,\ms\ RMS (i.e., comparable to the estimated RV noise of $\sigma$\,=\,5\,\ms). Modeling these RVs using the method described in Sec.~\ref{sec:sec3_2} unsurprisingly yields consistent planet and stellar activity parameters with those obtained in Sec.~\ref{sec:rv_an} (\kp\,=\,10.1\,$\pm$\,2.3\,\ms\ when $S$ is assumed to be null, and
\kp\,=\,10.1$^{+3.1}_{-2.9}$\,\ms\ otherwise, with $S$\,=\,6.4$^{+3.2}_{-2.9}$\,\ms; see lines 8 and 9 of Tab.~\ref{tab:results_rv}). This confirms that our final set $\boldsymbol{I_{\rm{f}}}$ of LSD profiles retains the original time-dependent content, and that the profile distortions caused by surface features were mostly unaffected by our procedure.

Following \citet{petit2015}, we reconstruct the surface brightness of the star with ZDI at the same time as the main planet orbital parameter, \kp. Moreover, we simultaneously include the modeling of surface differential rotation (DR) as described in \citet{donati2000} and \citet{petit2002}, to ensure a fully consistent analysis of our observations. In practice, DR is assumed to be solar like, with the rotation rate $\Omega(\theta)$ at the surface of the star, given by


\begin{eqnarray}
\Omega (\theta) = \Omega_{\rm{eq}} - \left( \cos{\theta} \right)^{2} d\Omega,
\label{eq:DR_formulae}
\end{eqnarray}

\noindent
where $\theta$ is the colatitude, \oeq, the equatorial rotation rate and, \dome, the difference in rotation rate between the pole and the equator. The planetary signal is modeled using Eq.~\ref{eq:rv_p} and assuming that \php\,=\,0, which leaves \kp\ as the only unknown planet parameter.

We carry out brightness reconstructions of $\boldsymbol{I_{\rm{f}}}$ for a wide range of DR and planet parameters. For each value of the parameters in the grid, (\kp, \oeq, \dome), the reduced Stokes $I$ LSD profiles are corrected from the barycentric motion induced by the planet using Eq.~\ref{eq:rv_p}, and, then, reconstructed with ZDI for a given amount of information at the surface of the star (i.e., a given amount of spot coverage). Following \citet{press1992}, we derive the best parameters by fitting a 3D paraboloid to the resulting \chs\ map around the minimum, and compute the 1$\sigma$ error bars on each parameter from the curvature of the modeled paraboloid. Note that the optimal parameters and the associated error bars no more than marginally depend on the level of spot coverage adopted for ZDI reconstructions.

We find that \oeq\,=\,1.298\,$\pm$\,0.003\,rad/d, \dome\,=\,0.075\,$\pm$\,0.031\,rad/d and \kp\,=\,9.7\,$\pm$\,2.5\,\ms\ provides the best reconstruction (see the 2D cuts shown in Fig.~\ref{fig:3D_DR}). The DR parameters are consistent with a solar-like differential rotation, with rotation periods of 4.84\,$\pm$\,0.01\,d at the equator and 5.10\,$\pm$\,0.15\,d at the pole (hence a difference of $\Delta$\pr\,=\,0.26\,$\pm$\,0.15\,d between the pole and the equator; note that the solid-body rotation is ruled out at 2$\sigma$). \kp\ is consistent within the error bars with the value obtained from the our GP modeling of Sec.~\ref{sec:sec3.3}. The reconstructed brightness map at the best planet and DR parameters is shown in Fig.~\ref{fig:surf_bright}. It features a spot coverage of 1.4\%, with almost equal proportions of dark and bright surface features divided into two distinct groups on each side of the star. Note that surface inhomogeneties other than brightness variations (e.g., small-scale magnetic fields), may also generate similar profile variations;  we come back on this point in Sec.~\ref{sec:conclusion}.

\begin{figure}
    \centering
    \includegraphics[width=\linewidth]{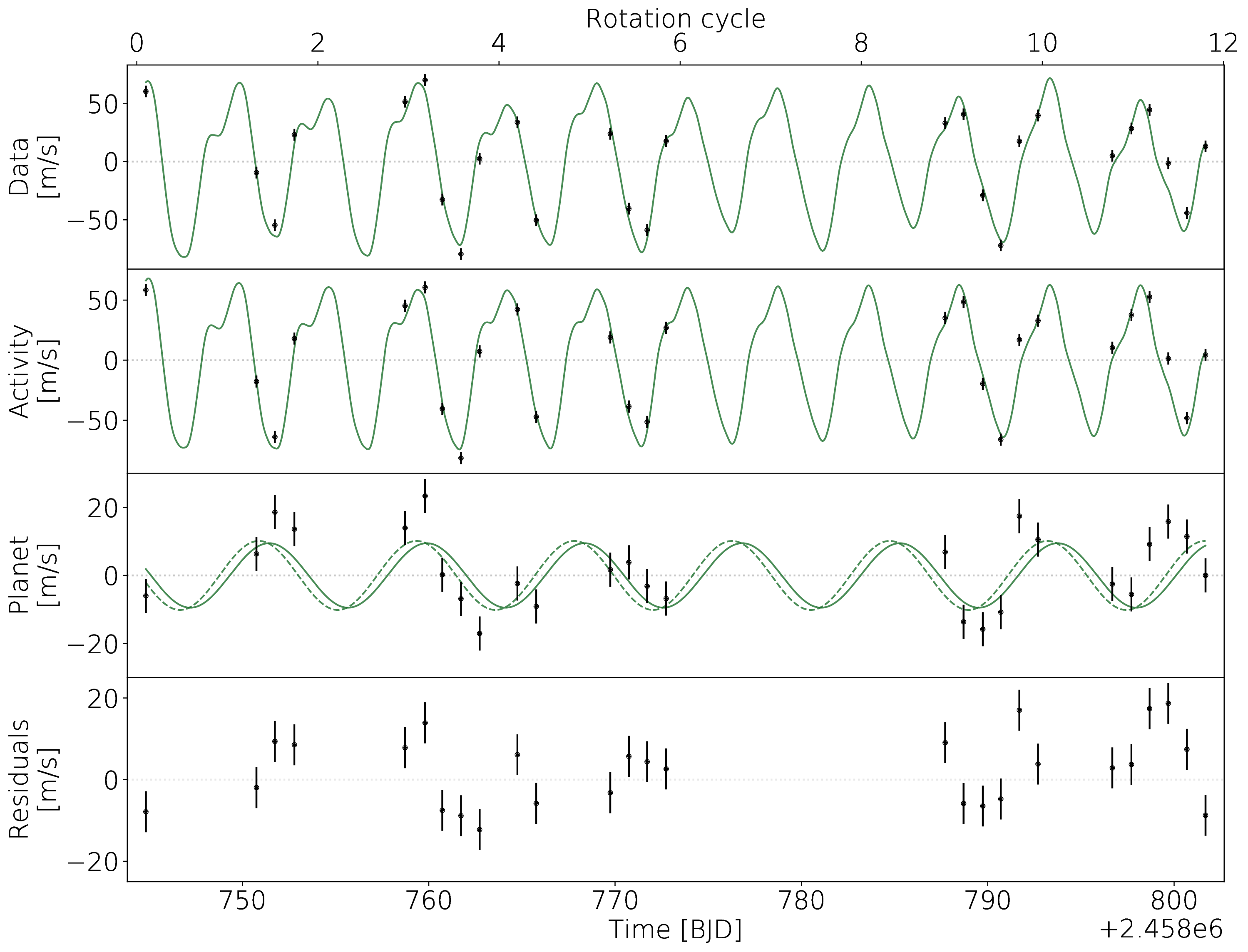}
    \caption{Same as Fig.~\ref{fig:pred_rv}, but, this time, using the maximum entropy brightness reconstruction of the star to generate the stellar activity RV curve. The solid and dashed lines in panel 3 show the recovered planet RV signatures when \php\,=\,0.0 and when \php\ is a free parameter of the model, respectively. The RMS of the residuals is 9\,\ms.}
    \label{fig:rv_zdi}
\end{figure}

To double-check that the planet signature evidenced through our 3D paraboloid fit is truly present in the RV data, we compute the stellar activity RV signal that corresponds to our optimum maximum entropy brightness map, and use it to filter stellar activity from our reference RV curve. To achieve this, we generate synthetic Stokes $I$ profiles (assuming \kp\,=\,0.0\,\ms\ and the best DR parameters obtained from the ZDI reconstruction), whose RVs are subtracted from the observed RV time-series (here, the RVs measured from $\boldsymbol{I_{\rm{f}}}$). We then fit the planetary model of Eq.~\ref{eq:rv_p} to the activity-subtracted RVs using a least-squares estimator. The best fit to the data is shown in Fig.~\ref{fig:rv_zdi}. The dispersion in the residuals is 9\,\ms\ RMS, i.e., $\sim$3$\times$ larger than when the stellar activity is modeled with GP. This is expected as GPs are, by nature, more flexible than the current version of ZDI, and thereby capable of fitting the temporal evolution of RV curves as a result of intrinsic variability, whereas ZDI can only model that caused by DR and not that from temporal evolution of surface features. Unsurprisingly, we obtain that \kp\,=\,9.5\,$\pm$\,2.5\,\ms, fully consistent with the value obtained from the 3D paraboloid fit.\footnote{To empirically take into account the excess dispersion in the RV residuals, error bars on \kp\ were scaled up by $\sqrt{}$\crr\,$\approx$\,9/5\,=\,1.8}.

\section{Magnetic analysis}\label{sec:mag_an}

\begin{figure*}
    \centering
    \includegraphics[width=0.49\linewidth]{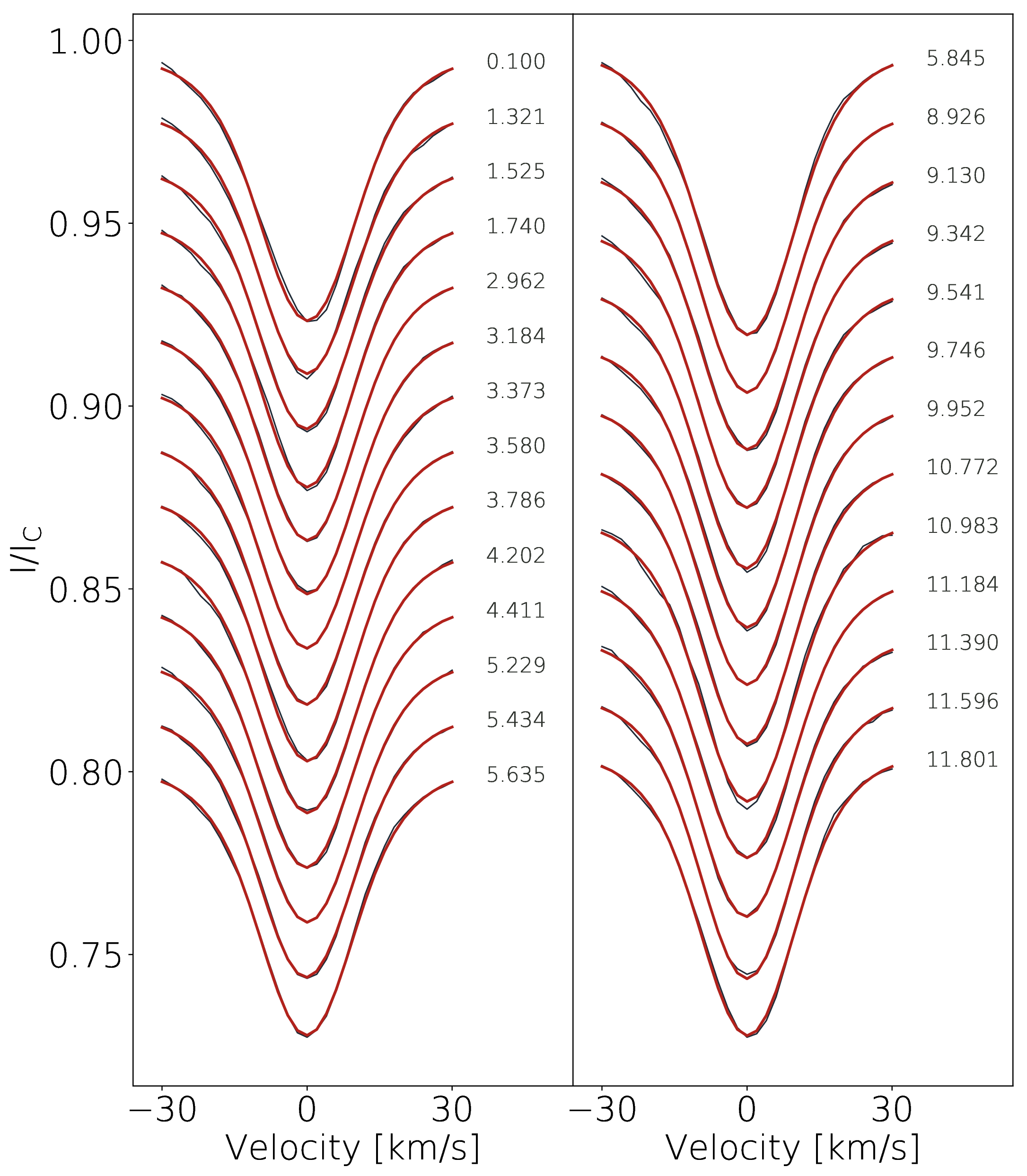}
    \includegraphics[width=0.49\linewidth]{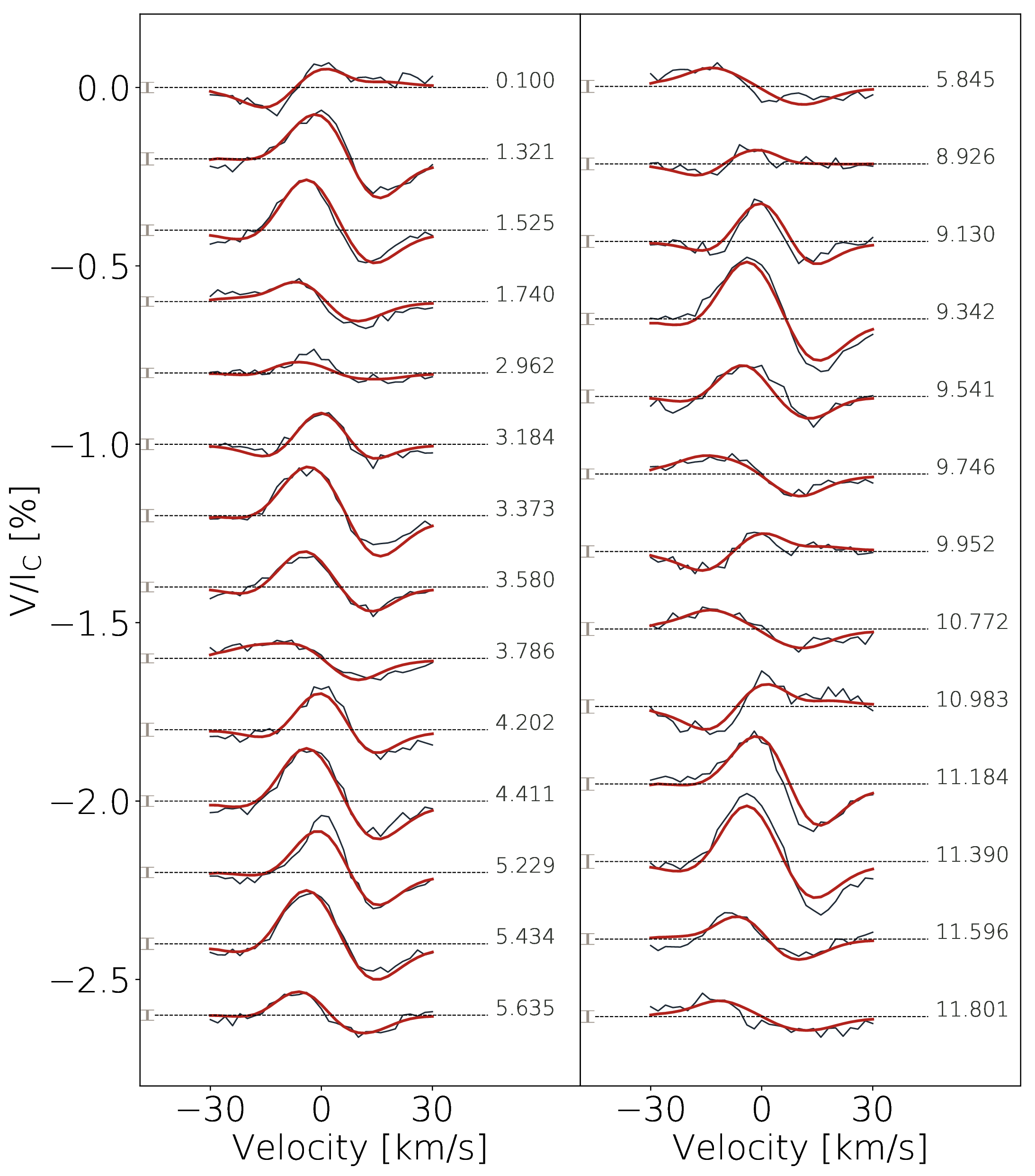}
    \caption{Best fit to the observed Stokes $I$ (left panel) and Stokes $V$ (right panel) LSD profiles. The figure elements are the same as in Fig.~\ref{fig:StokesI_Q}. The $\pm$\,1$\sigma$ error bars are indicated to the left of each Stokes $V$ profile.}
    \label{fig:fit_stokes}
\end{figure*}


We use ZDI to perform simultaneous maximum-entropy reconstructions of the relative brightness and large-scale magnetic field at the surface of \aum\ from the observed Stokes $I$ and Stokes $V$ LSD profiles, extracted using the mask of atomic lines presented in Sec.~\ref{sec:data_red}. ZDI decomposes the large-scale magnetic field vector into its toroidal and poloidal components, expressed as weighted sums of spherical harmonics \citep{donati2006}. The reconstruction process is similar to that described in Sec.~\ref{sec:bright_im}, except that we now also compute synthetic Stokes $V$ profiles, using again the Unno–Rachkovsky’s analytical solution to the polarized radiative transfer equations, which takes both local brightness and magnetic field into account. In order to deal with the strong broadening of the Zeeman signatures in the observed Stokes $I$ and $V$ LSD profiles, we assume that a fraction $f$ of each cell (constant over the star) includes magnetic fields of intensity $B/f$, whereas the rest of the cell is free of magnetic field, so that the large-scale magnetic flux over the cell is equal to $B$.


Using the same stellar parameters as for ZDI, we first invert the Stokes $V$ time-series alone, pushing the spherical harmonic expansion to degree $l$\,=\,7, given the relatively low \vs\ of the star \citep[e.g.,][]{morin2008}. Assuming solid-body rotation, we find that \pr\,=\,4.80\,$\pm$0.01\,d minimizes the \crr\ of the reconstruction, lower than the photometric rotation period of \aum\ reported in \citetalias{plavchan2020}. We then perform combined brightness and magnetic inversions of the observed LSD profiles and find that $f$\,=\,0.2 reproduces best the Stokes $I$ and Stokes $V$ time-series down to a unit \crr\ for both Stokes $I$ and Stokes $V$ profiles, starting from values of 2.8 and 11.8 respectively.

\begin{figure*}
    \centering
    \includegraphics[width=\linewidth]{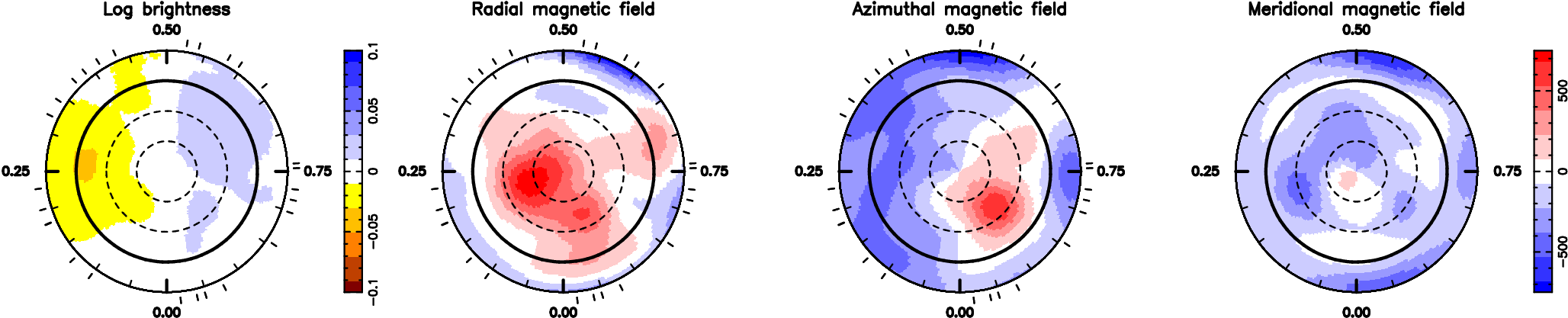}
    \caption{Logarithmic brightness (left panel) and radial, azimuthal and meridional components of the large-scale magnetic field (resp. panels 2 to 4) at the surface of \aum. Each panel is a flattened polar view featuring the same symbolds and notations as Fig.~\ref{fig:surf_bright}. Magnetic fluxes are expressed in G.}
    \label{fig:zdi_map}
\end{figure*}

The maximum-entropy fit to the observed Stokes $I$ and $V$ LSD profiles is given in Fig.~\ref{fig:fit_stokes} and the corresponding brightness and large-scale magnetic maps are shown in Fig.~\ref{fig:zdi_map}. As expected, the recovered brightness map is similar to that reconstructed from our previous series of Stokes $I$ LSD profiles (see Sec.~\ref{sec:bright_im} and Fig.~\ref{fig:surf_bright}). For the magnetic map, we find an average large-scale magnetic flux at the surface of the star of 475\,G, and featuring a dipole of 450\,G tilted at 19\degr\ to the rotation axis towards phase 0.18. The reconstructed field is found to be mainly poloidal and axisymmetric, with the poloidal component enclosing 78\% of the derived magnetic energy, 65\% of which in axisymmetric modes. The filling factor we obtain ($f$\,=\,0.2) means that a magnetic field of $\sim$2.4\,kG is present on $\sim$20\% of the stellar surface, consistent with the average strength of the magnetic field measured from the Zeeman broadening of unpolarized lines \citep{saar1994,reiners2012,shulyak2017,moutou2017}. Although the dipole is the dominant component of the reconstruction, we find that quadrupolar, octupolar and smaller scale structures enclose a significant fraction of the total magnetic energy budget (resp. 8, 9 and 15\%).

\begin{figure}
    \centering
    \includegraphics[width=\linewidth]{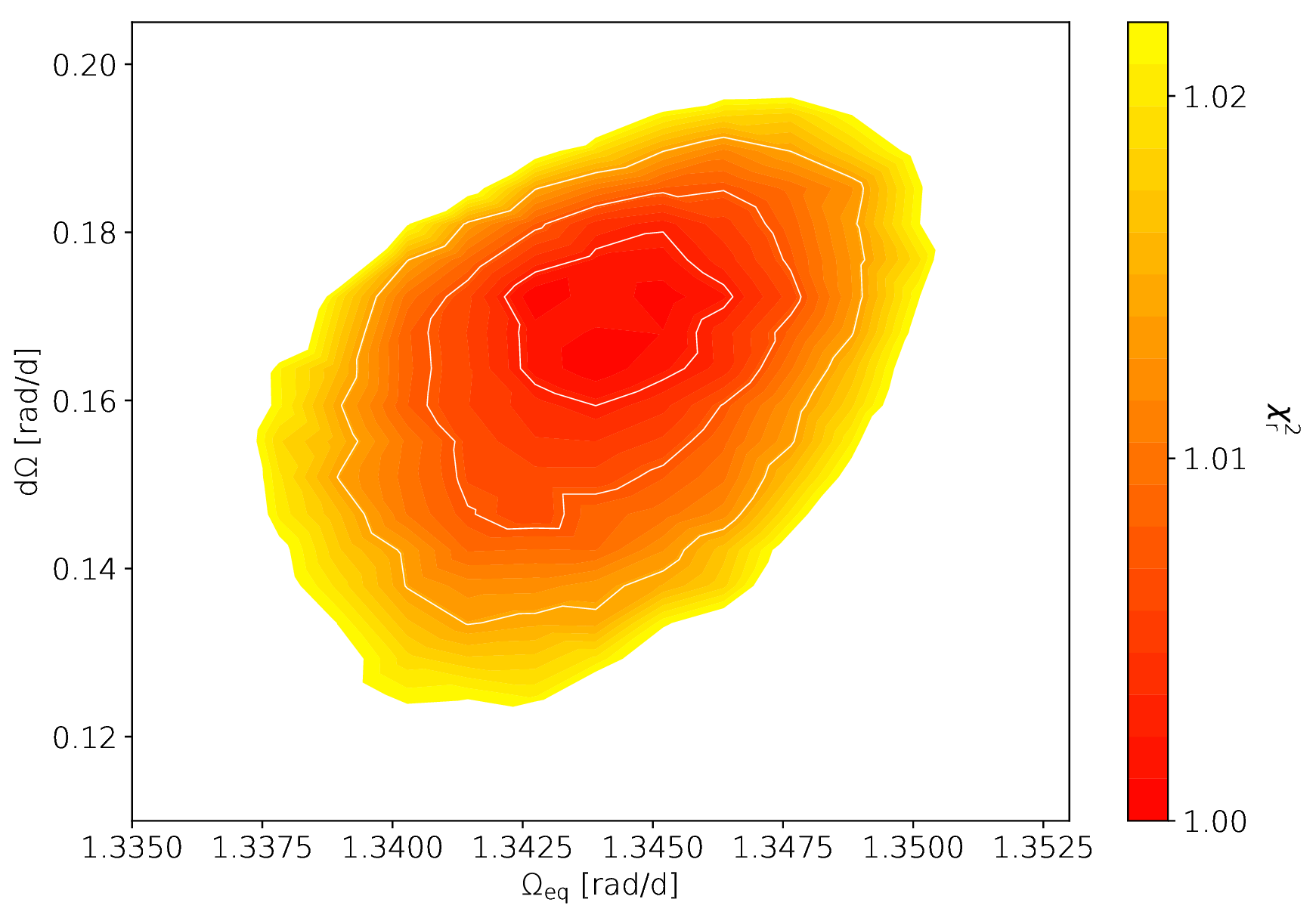}
    \caption{\crr\ as a function of the DR parameters \oeq\ and \dome\ derived from ZDI modeling of our time-series of Stokes $V$ LSD profiles. Similarly to Fig.~\ref{fig:3D_DR}, white solid lines indicate 1, 2 and 3$\sigma$ contours around the minimum \crr.}
    \label{fig:dr_map_V}
\end{figure}

From the time-series of Stokes $V$ LSD profiles alone, we estimate the stellar DR using the method introduced in Sec.~\ref{sec:bright_im}. ZDI reconstructions are carried out for a range of (\oeq, \dome), leading to the 2D \crr\ map shown in Fig.~\ref{fig:dr_map_V}, from which we estimate the DR parameters with error bars using a 2D-paraboloid fit \citep[see][]{donati2003}. We find that \oeq\,=\,1.344\,$\pm$\,0.002\,rad/d and \dome\,=\,0.167\,$\pm$\,0.009\,rad/d minimizes the magnetic information at the surface of \aum. This implies that the rotation period of the large-scale field varies from 4.675\,$\pm$\,0.006\,d, around the equator, to 5.34\,$\pm$\,0.05\,d, at the pole (hence a difference of $\Delta$\pr\,=\,0.66\,$\pm$\,0.05\,d).



\section{Activity indicators}\label{sec:act_ind}

\begin{figure}
    \centering
    \includegraphics[width=\linewidth]{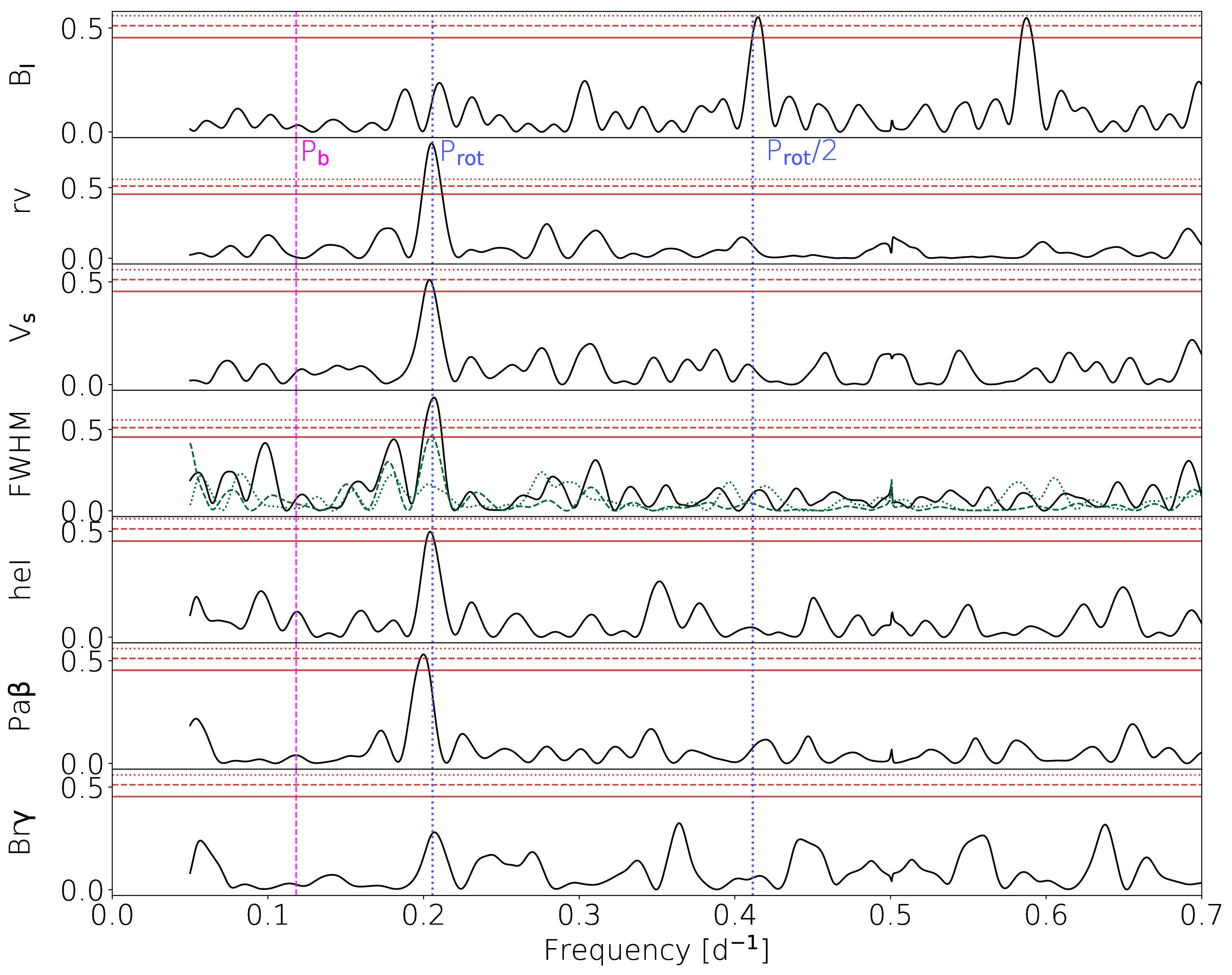}
    \caption{From top to bottom: GLS periodograms of \bl, RV, \vs, FWHM, He\,I, Pa$\beta$ and Br$\gamma$. Symbols are as for Fig.~\ref{fig:period_rv}, except for FAPs that are computed at 0.3, 0.1 and 0.03. In panel~4, the green dashed and dotted lines show the GLS periodograms of the time series of FWHMs of the profiles computed with the atomic and empirical line masks described in Sec.~\ref{sec:data_red} (and respectively corresponding to the Stokes $I$ profiles shown in Fig.~\ref{fig:StokesI_Q} and Fig.~\ref{fig:fit_stokes}).}
    \label{fig:period_ind}
\end{figure}

Optical chromospheric lines such as H$\alpha$- and Ca\,H\&K are well-known proxies of stellar activity RV signals \citep[e.g.,][]{bonfils2007,boisse2009,gomes2011} and correlate generally well with the topology of large-scale magnetic field at the stellar surface \citep[e.g.,][]{hebrard2016}. In the nIR, only a couple of chromospheric lines have been identified as potential probes of stellar activity at the photospheric level \citep[e.g., He\,I 1083\,nm and Pa$\beta$;][]{zirin1982,short1998}, even if the way they are connected to the photosphere or transition region in low-mass stars remains unclear \citep[e.g.,][]{sanz-forcada2008,schmidt2012,schofer2019,moutou2020}. Here, we propose to compute indices based on He\,I triplet (1083\,nm), Pa$\beta$ (1282\,nm) and Br$\gamma$ (2165\,nm). Each telluric-corrected Stokes $I$ spectrum is divided by the median spectrum, in the stellar rest frame. The activity indices are computed by taking the average of each median-divided spectrum in intervals centered on the lines of interest\footnote{Each chromospheric activity index defined here is thus equal to $1 + \eta$, where $\eta$\,=\,$\delta f_{\rm{I}}$/$f_{\rm{c}}$, where $\delta f_{\rm{I}}$ is the difference of flux relative to the median value, and, $f_{\rm{c}}$, the flux of the continuum in the considered wavelength domain.} (i.e., between 1082.82 and 1083.1\,nm for He\,I, 1281.72 and 1281.9 for Pa$\beta$, and 2165.32 and 2165.72 for Br$\gamma$). For each observation the error bars are computed from the average dispersion within 0.5\,nm wide windows at each side of the lines of interest (see the values listed in Tab.~\ref{tab:App_A}). No flare is detected in the time-series of chromospheric indicators.


The small-scale magnetic flux has been shown to be an excellent proxy of stellar activity RV signals in the case of the Sun \citep{haywood2016,haywood2020}. The temporal changes of this quantity are closely linked to those of the broadening of Zeeman-sensitive lines. In a way similar to that described in Sec.~\ref{sec:data_red}, we apply LSD to our reduced set of spectra but, this time, with a mask including the $\sim$700~lines with Land\'e factors larger than 1.5. We then measure the FWHM of each profile by modeling it by the sum of a Gaussian function and a linear continuum slope. In what follows, we refer to this indicator as FWHM.




The GLS peridograms of the \bl, RV, velocity span ($V_{\rm{S}}$), FWHM, He\,I, Pa$\beta$ and Br$\gamma$ time-series are presented in Fig.~\ref{fig:period_ind}. Except for Br$\gamma$, whose rotational modulation is marginally detected, all activity indicators exhibit a prominent peak around \pr\ or its first harmonic. In particular, we note that the rotational modulation in the FWHM of Stokes $I$ LSD profiles increases for lines with increasing magnetic sensitivity (i.e., increasing Land\'e factors). For example, the profiles computed from our empirical mask, which includes a majority of molecular lines weakly sensitive to magnetic fields, shows no more than a weak modulation.


We use GP with the quasi-periodic kernel of Eq.~\ref{eq:cov_fct} to independently model \bl, $V_{\rm{S}}$, FWHM, as well as He\,I- and Pa$\beta$-based indices. For the modeling of the \bl\ time-series, all hyperparameters are optimized by the MCMC process with Jeffreys and uniform prior densities for $\theta_{2}$ and $\theta_{4}$, respectively. For the other time-series, we fix $\theta_{2}$ to 100\,d, while $\theta_{4}$ is set to 1.0 for chromospheric indicators and FWHM, and optimized for the $V_{\rm{S}}$ time-series\footnote{$\theta_{4}$ is set to 1.0 when modeling the chromospheric activity indicators, whereas 0.4 was used in the case of the RVs. This is because active regions at the stellar surface produce RV signatures evolving about twice faster with time than their counterparts in the curves of activity proxies. The recurrence time-scale of the GP ($\theta_{3}$) is found to be marginally affected by variations of the order of 0.1 in $\theta_{4}$.}. As for the RV analysis, we included a constant offset and an excess uncorrelated noise to the model. The resulting rotation periods and error bars are given in Tab.~\ref{tab:prot_indic} (all the results of the fits are given in Tab.~\ref{tab:fit_indic}, while the best predictions are shown in Fig.~\ref{fig:pred_GP_ind}).


We find that \pr\,=\,4.83\,$\pm$\,0.02\,d describes best the rotational modulation of the \bl\ time-series. This value is consistent with that obtained from the RV analysis, confirming that \bl\ is a reliable proxy of stellar rotation periods \citep{donati2006,hebrard2016}. The same applies to $V_{\rm{S}}$ and FWHM, which appear modulated at rotation periods of 4.89$^{+0.04}_{-0.02}$\,d and 4.84\,$\pm$\,0.04\,d, respectively. The modeling of the chromospheric indices yields respective periods of 4.87$^{+0.06}_{-0.05}$ and 4.99$^{+0.11}_{-0.09}$\,d for He\,I and Pa$\beta$, both compatible with the rotation period of the star.



{\renewcommand{\arraystretch}{1.25} 
\begin{table*}
    \centering
    \caption{Best rotation period (line~1) from GP modeling of the different activity indicators described in Sec.~\ref{sec:act_ind}, and Pearson correlation coefficient ($\rho$, line~3) of these indicators with the planet-subtracted RVs. The last column indicates the rotation period obtained from the ZDI modeling of the Stokes $V$ LSD profiles (see Sec.~\ref{sec:mag_an}). For comparison, the DR parameters estimated from the time series of Stokes $I$ and Stokes $V$ profiles yield respective rotation periods of 4.84\,$\pm$\,0.01\,d and 4.675\,$\pm$\,0.006\,d at the stellar equator, and 5.10\,$\pm$\,0.15\,d and 5.34\,$\pm$\,0.04\,d at the stellar pole.}
    \label{tab:prot_indic}
    \begin{tabular}{cccccccc}
        \hline
        & RVs & \bl & $V_{\rm{S}}$ & FWHM & He\,I & Pa$\beta$ & Stokes $V$ \\
        \hline
        \pr & 4.84\,$\pm$\,0.01\,d & 4.83\,$\pm$\,0.02 & 4.89$^{+0.04}_{-0.02}$ & 4.84\,$\pm$\,0.04 & 4.87$^{+0.06}_{-0.05}$ & 4.99$^{+0.11}_{-0.09}$ & 4.80\,$\pm$\,0.01  \\
        $\rho$ & -- & -0.214 & -0.695 & 0.691 & -0.503 & -0.281 & -- \\
        \hline
    \end{tabular}
\end{table*}}

\begin{figure}
    \centering
    \includegraphics[width=\linewidth]{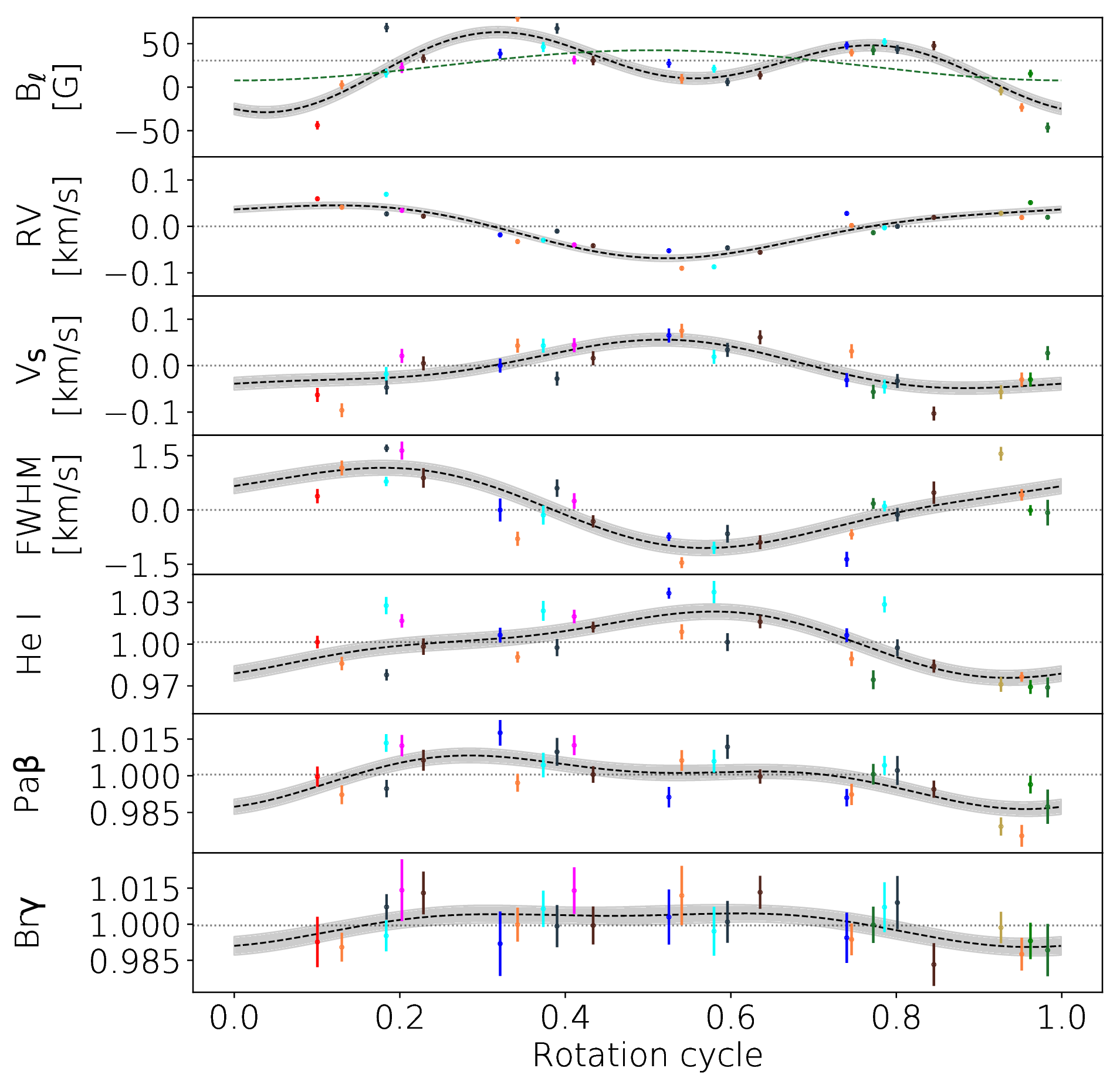}
    \caption{From top to bottom: time series of \bl\, RV, $V_{\rm{S}}$, FWHM, He\,I, Pa$\beta$ and Br$\gamma$, phase-folded at a rotation period of 4.86\,d. In each panel, a double sine-wave fit (with 1 harmonic) to the time-series assuming a period of \pr\ is shown in black dashed lines (with $\pm$1$\sigma$ error bands). Similarly to Fig.~\ref{fig:phase_fold_rv}, data points of different colors belong to different rotational cycles. In the top panel, we show the best simple sine-wave fit to the \bl\ time-series at a rotation period of 4.86\,d.}
    \label{fig:ph_fold_ind}
\end{figure}

The time-series of activity indicators are phase-folded at \pr\,=\,4.86\,d in Fig.~\ref{fig:ph_fold_ind}. At the first order, the chromospheric indicators appear well-correlated with \bl\ and anti-correlated with RV. In particular, the longitudinal field appears best-correlated with Pa$\beta$, with a maximum around phase 0.3, when the magnetic pole is pointing toward the observer. In contrast, the He\,I flux is maximum around phase 0.6, when the magnetic equator faces the observer (the magnetic dipole is tilted at 19\degr\ to the rotation axis; see Fig.~\ref{fig:zdi_map}), emphasizing the fact that He\,I is mostly observed at low latitudes. We also find that the FWHM is remarkably phased with RVs, suggesting that the small-scale magnetic field and the regions of the stellar disk responsible for the RV fluctuations are closely located.

\begin{figure*}
    \centering
    \includegraphics[width=\linewidth]{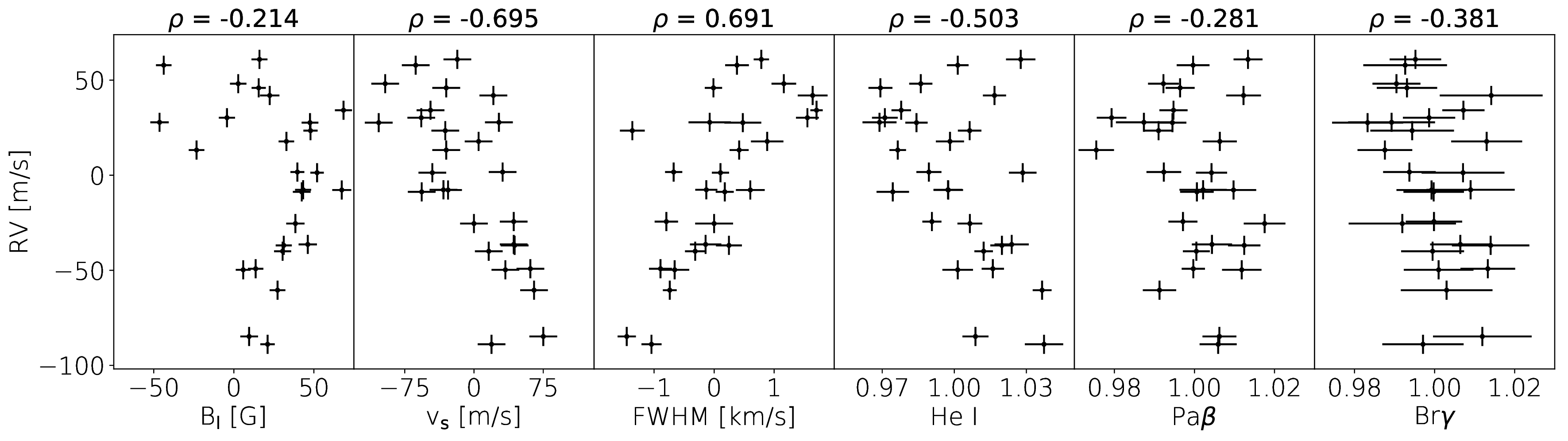}
    \caption{Correlation plots of the RV time-series (measured by fitting a Gaussian function to each Stokes $I$ LSD profile) as a function of (from left to right): the longitudinal field, velocity span, FWHM, He\,I, Pa$\beta$ and Br$\gamma$. The Pearson correlation coefficient, $\rho$, is written at the top of each panel. Note that the planet signature was removed from the RVs for these curves, and before computing the correlation coefficients.}
    \label{fig:correl_ind}
\end{figure*}

In order to compare the ability of the indicators to model stellar activity RV signals, we compute the Pearson correlation coefficient, $\rho$, between each time series of indicators and the planet-subtracted RVs measured in Sec.~\ref{sec:sec3_1}. The resulting correlation plots are shown in Fig.~\ref{fig:correl_ind} (see also all the correlations in Fig.~\ref{fig:ind_correl_all}). The RV time-series is best correlated with $V_{\rm{S}}$ and FWHM, confirming that both the velocity span and the small-scale magnetic flux are reliable proxies of stellar activity for moderate-to-fast rotators in the nIR. We also observe a moderate anti-correlation between RVs and He\,I, suggesting that this indicator describes relatively well the variations induced by non-axisymmetric features on the line profiles. Although we find that our RVs are well-phased with a simple sine fit to the \bl\ time-series (see the top panel of Fig.~\ref{fig:ph_fold_ind}), higher frequency structures in \bl\ reduce the correlation with the RVs. No significant correlation is identified between RVs and Pa$\beta$, that appear more sensitive to magnetic structures located at higher latitudes. Finally, Br$\gamma$ has no visible modulation and appears to be a poor indicator of stellar activity in the case of \aum.

\section{Discussion and conclusion}\label{sec:conclusion}

In this study, we conducted a velocimetric and spectropolarimetric analysis of a data set consisting of 27 observations of \aum\ collected with SPIRou from 2019 September 18 to November 14. Using Least-Squares Deconvolution, we computed average Stokes $I$ and Stokes $I$ and $V$ profiles using 2 different line masks, one empirically built from a SPIRou spectrum of an early M dwarf (Gl\,15A) and containing both atomic and molecular lines, and another including atomic lines with known Land\'e factors only. Stellar RVs are measured with a conservative 1$\sigma$-uncertainty of 5\,\ms, and exhibit fluctuations of up to 150\,\ms\ peak-to-peak with a RMS of 45\,\ms. A third mask including lines with strongest magnetic sensitivity was also used to estimate the amount of Zeeman broadening of Stokes $I$ profiles, and its modulation with time.

We carried out the search for a planet signature at \aum~b's orbital period, assuming a circular orbit, in our RV time-series. Stellar activity and planet RV signals were jointly estimated, resulting in a 3.9$\sigma$ detection of the planet-induced RV signature, implying a mass of 17.1$^{+4.7}_{-4.5}$\,\mpear\ (i.e., 1.0\,$\pm$\,0.3\,$M_{\neptune}$). Including an excess of uncorrelated noise $S$ in the free parameters of the model yields a planet mass estimate of 18.7$^{+6.5}_{-6.1}$\,\mpear, compatible with the previous value but with a larger uncertainty. However, simulations demonstrate that the value of $S$ recovered from the fit to the data is most likely over-estimated in our limited data set, leading to artificially enhanced error bars on the planet mass. We thus consider that the planet mass derived assuming $S$\,=\,0\,\ms\ is the most reliable and robust estimate that one can retrieve from our data set.

A consistent measurement of the planet mass is independently obtained by performing a reconstruction of the surface brightness of the star with ZDI while estimating the planet parameters. In this process, the stellar activity RV signal is modeled by bright and dark features at the surface of the star. In fact, it is entirely possible that the reconstructed surface brightness inhomogeneities modulating the width of Stokes $I$ profiles probe (at least partly) small-scale magnetic fields rather than brightness features. The fact that the FWHM of lines with higher magnetic sensitivity is more strongly modulated (see Fig.~\ref{fig:period_ind}) is further evidence that the activity-induced RV signal is indeed of magnetic origin. Breaking the degeneracy between surface brightness and magnetic field would require to simultaneously model Stokes I LSD profiles of spectral lines with different magnetic sensitivities, so that the differences in their modulation pattern can be reliably inverted into maps of the surface brightness and small-scale magnetic field. As this approach requires a full-size study in itself, we postpone it to a forthcoming paper.



\begin{figure}
    \centering
    \includegraphics[width=\linewidth]{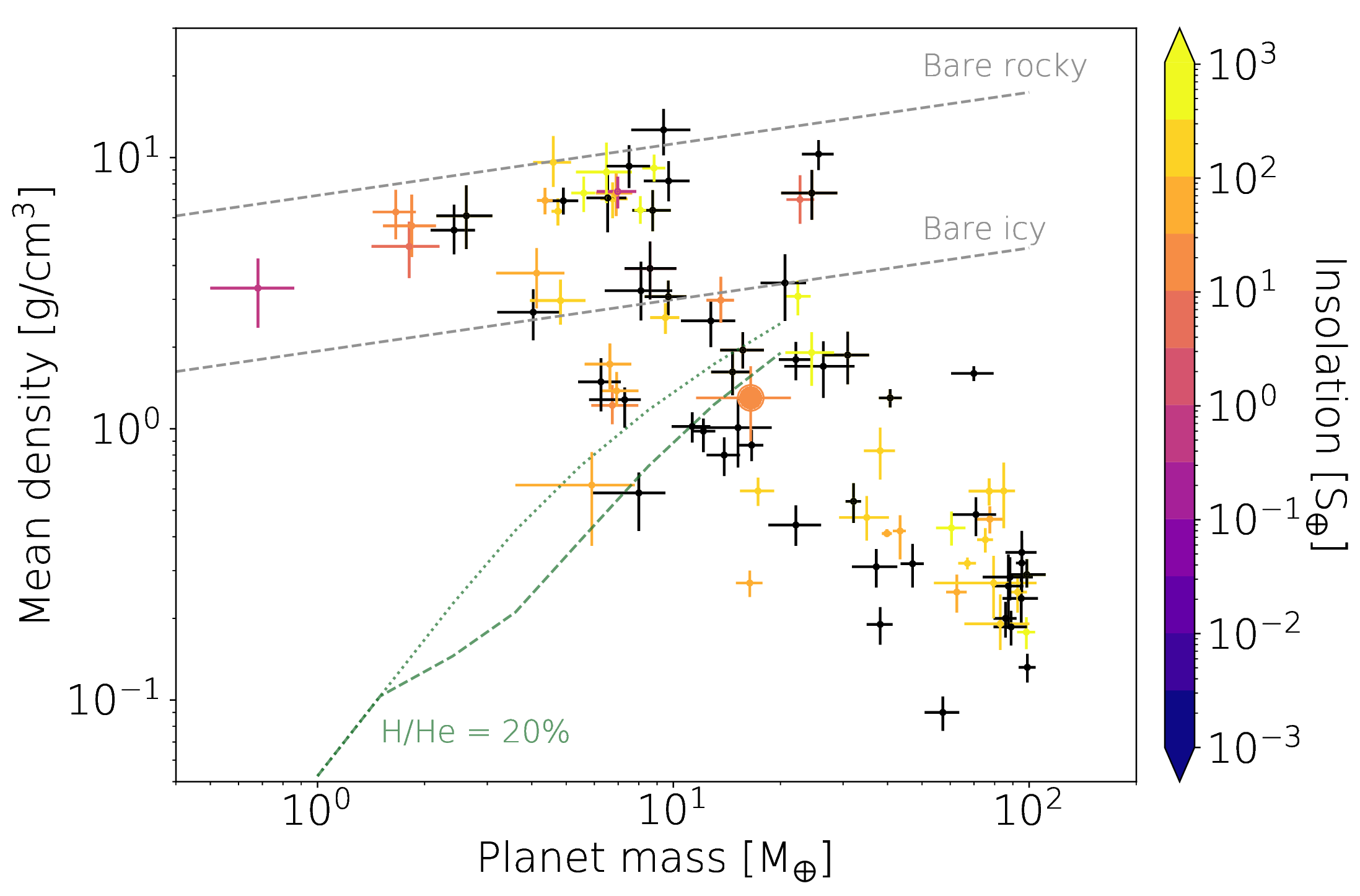}
    \caption{Mass-mean density diagram of confirmed exoplanets of mass lower than 100\,\mpear\ and with relaive uncertainties on the mass and the mean density lower than 33\%, computed from \url{https://exoplanetarchive.ipac.caltech.edu/}. The color scale depicts the insolation of the planet (planets are plotted in black when insolation is not known). \aum~b is indicated by the large filled circle. For comparison purposes, we plot the theoretical limits of pure icy and rocky cores (i.e., completely evaporated envelope), computed using \citet{jin2018}'s relations. Green dashed and dotted lines indicate the expected density of a 100\,Myr planet under \aum~b's stellar flux hosting a 20\% H/He envelope assuming respectively solar and 50$\times$ solar planet metallicities predicted from \citet{lopez2014}'s theoretical models. Note that all planets but \aum~b shown in this diagram orbit stars older than 100\,Myr.}
    \label{fig:pl_diagram}
\end{figure}

When combined with the transit radius, the planet mass yields a mean density of 1.3\,$\pm$\,0.4\,g\,cm$^{-3}$, comparable to that of Neptune within the 1$\sigma$ error bars, which is relatively high compared to the prediction of global models of planet formation and evolution, and suggests that the planet features a high fraction of heavy elements \citep[typically $\ga$\,80\%; e.g.,][]{mordasini2012b}. \aum~b is compared to older low-mass exoplanets with well-constrained mean density in the mass-mean density diagram shown in Fig.~\ref{fig:pl_diagram}. When compared to the theoretical evolution of the mass-radius relation of sub-Neptunes of \citet{lopez2014}, \aum~b is expected to be a core-dominated planet featuring a H/He envelope occupying \ga\,20\% of the total planet mass. We also note that \aum~b lies fairly close to the lower limit of the so-called evaporation valley \citep[e.g.,][]{owen2013,lopez2013,jin2014,jin2018,mordasini2020}. \aum~b has an equilibrium temperature ($\sim$600\,K) and radius of the same order of the well-known evaporating sub-Neptune GJ\,436~b \citep[e.g.][]{bourrier2018} and may thus be in the process of loosing its H/He envelope which would lead it to cross the evaporation valley on time scales of a few tens of Myr \citep{jin2018}. Measuring the metallicity of the planet atmosphere, as well as detecting extended exospheres of hydrogen or helium, will greatly improve our understanding of the nature of \aum~b, its formation history, and its future evolution.

\begin{table*}
    \centering
    \caption{Comparison of \aum\ and AD\,Leo stellar properties. Lines~7 and~8 give respectively the relative depth of the convective envelope, $R_{\rm{C}}$ (in unit of stellar radius, \rs), and the moment of inertia relative to $M_{\rm{S}}$\rs$^{2}$, $k_{\rm{i}}$, computed from \citet{siess2000} and \citet{baraffe2015} models. The last 4 lines give the average intensity of the large-scale magnetic field, $B_{\rm{V}}$, and the fractions of poloidal ($f_{\rm{pol}}$), dipolar (poloidal, $f_{\rm{dip}}$), and axisymmetric (poloidal, $f_{\rm{axi}}$) modes of the magnetic energy reconstructed by ZDI \citep[the two values given for AD\,Leo correspond respectively to the reconstructions of][]{morin2008,lavail2018}. The properties of \aum\ predicted at an age of 100\,Myr by \citet{siess2000}/\citet{baraffe2015} models, assuming the conservation angular momentum, are given in column~3. Except for lines~7 and~8, the references listed in column~5 refer to AD\,Leo (see Tab.~\ref{tab:star_prop} for the references associated to \aum\ properties). We note that the measured mass and age of AD\,Leo are in better agreement with the models of \citet{baraffe2015} than with those of \citet{siess2000}.}
    \label{tab:star_mag_prop}    
    \begin{tabular}{ccccc}
        \hline
        Parameter & \aum &  \aum\ at 100\,Myr   &  AD\,Leo & References for AD\,Leo \\
        \hline
        Age [Myr]  & 22\,$\pm$\,3 & 100  & 25-300 & \cite{shkolnik2009} \\
        \teff\ [K] & 3700\,$\pm$\,100 & 3750/3880  & 3471\,$\pm$\,60 & \cite{gaidos} \\
        Radius [\rsun] &  0.75\,$\pm$\,0.03  & 0.39/0.53 &  0.402\,$\pm$\,0.010 & \cite{mann2015} \\
        Mass [\msun] & 0.50\,$\pm$\,0.03 &  0.50  & 0.427\,$\pm$\,0.011 & \citet{mann2015} \\
        Luminosity [$L_{\odot}$] & 0.09\,$\pm$\,0.02 & 0.04/0.06 & 0.022\,$\pm$\,0.002 & From \teff\ and mass \\
        \pr\ [d] & 4.836\,$\pm$\,0.008 & 1.3/2.4  & 2.2399\,$\pm$\,0.0006 & \cite{morin2008}, \citet{tuomi2018} \\
        $R_{\rm{C}}$ [\rs] & 1.0/0.6 &  0.91/0.36 &  1.0/0.6 &  \cite{siess2000}/\cite{baraffe2015} \\
        $k_{\rm{i}}$ & 0.2/0.2 & 0.2/0.2  &  0.2/0.2 & \cite{siess2000}/\cite{baraffe2015} \\
        $B_{\rm{V}}$ [G] & 475 & -- & 190-300 &  \cite{morin2008}--\cite{lavail2018} \\
        $f_{\rm{pol}}$ [\%] & 80 & -- & 99-90 & \cite{morin2008}--\cite{lavail2018} \\
        $f_{\rm{dip}}$ [\%] & 60 & -- & 60-90 &  \cite{morin2008}--\cite{lavail2018} \\
        $f_{\rm{axi}}$ [\%] & 65 & -- & 90-95 &  \cite{morin2008}--\cite{lavail2018} \\
        \hline
    \end{tabular}
\end{table*}


Using ZDI, we derived the surface brightness and large-scale magnetic maps of \aum. Cool and warm features are present in roughly equal proportions, totalling 1-2\% of the whole stellar surface probed by ZDI. In particular, regions dominated by dark spots and warm plages are well-separated at the surface of the star. We find a large-scale magnetic field of 475\,G featuring a mainly poloidal and axisymmetric field dominated by a 450\,G-dipole component tilted at 19\degr\ to the rotation axis. We find that the large-scale field is sheared by a surprisingly large solar-like differential rotation of \dome\,=\,0.167\,rad/d (i.e., the equator rotates faster than the pole), about twice stronger than that observed in our brightness reconstruction and 3$\times$ larger than that of the Sun. The discrepancy between the equatorial periods derived from the Stokes $I$ and Stokes $V$ profiles suggests that both methods probe different layers of the convective zone. More specifically, by combining our DR parameters, we find that \oeq\,=\,0.48\dome\,$+$\,$\Omega_{0}$, where $\Omega_{0}$ is a constant (for the Sun, \oeq\,=\,0.2\dome\,$+$\,$\Omega_{0}$). This is compatible with the typical angular velocity fields of rapid rotators like AB\,Dor, best described with an angular rotation constant over axisymmetric cylinders \citep{donati2003}. In this context, the magnetic field would probe layers close to the surface while dark and bright features (and the related small-scale field) are likely anchored deeper in the convective zone. We caution that this result may be impacted by the high level of temporal variability of \aum's activity and, therefore, needs to be validated by additional spectropolarimetric observations.





Activity indicators based on \bl, $V_{\rm{S}}$, He\,I, Pa$\beta$, Br$\gamma$ and FWHM (probing the small-scale field) were computed at each observing epoch. Except for Br$\gamma$, all time series exhibit a clear rotational modulation. Interestingly, time series associated with He\,I and Pa$\beta$ are modulated with different periods, suggesting that they probe different regions of the stellar surface, He\,I being mostly coupled with the magnetic equator, while Pa$\beta$ is more sensitive to the magnetic pole and, as such, is more correlated with \bl\ than with RVs. Since the non-axisymmetric features reconstructed with ZDI are mostly located around the stellar equator, He\,I stands for a better proxy of the stellar activity RV signal than \bl\ or Pa$\beta$. Finally, we note that both the bisector span and the small-scale magnetic field correlates best with our RV time-series, confirming their high sensitivity to stellar activity, even in the nIR.

\begin{figure}
    \centering
    \includegraphics[width=\linewidth]{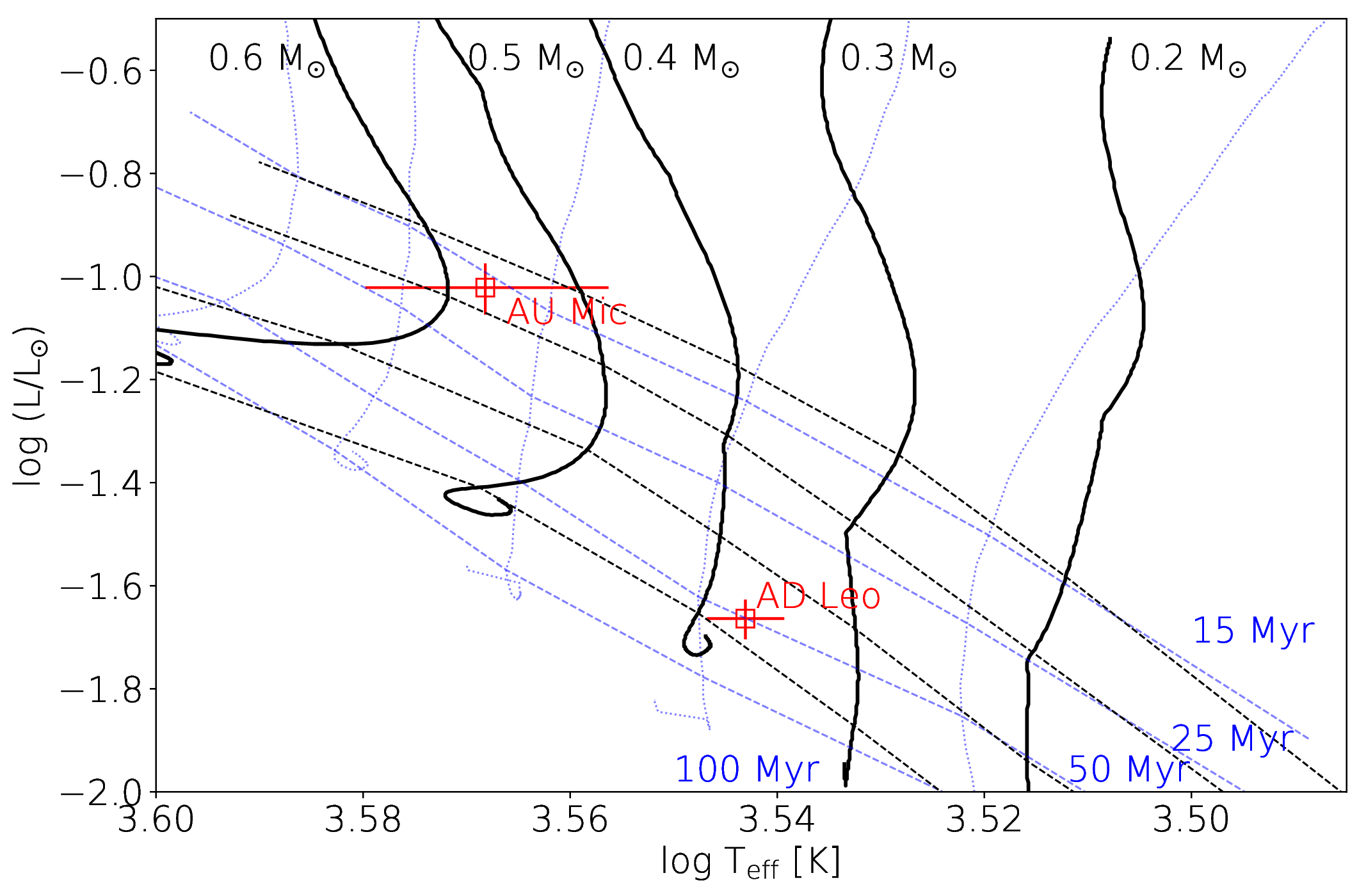}
    \caption{Location of \aum, and AD\,Leo in the HR diagram (with 1$\sigma$ error bars). The PMS evolutionary tracks for masses of 0.2, 0.3, 0.4, 0.5 and 0.6\,\msun\ are plotted in thick black solid lines \citep[for the models of][]{baraffe2015} and thin blue dotted lines \citep[for the models of ][assuming solar metallicity and including convective overshooting]{siess2000}. The corresponding isochrones at 15, 25, 50 and 100\,Myr are plotted in black and blue dashed lines.}
    \label{fig:star_diagram}
\end{figure}

The magnetic properties of \aum\ are relatively similar to those of the more evolved star AD\,Leo (see the compared stellar properties of Tab.~\ref{tab:star_mag_prop}). Both stars are located on close evolutionary tracks in the HR diagram (see Fig.~\ref{fig:star_diagram}), and one may wonder if \aum\ will resemble AD\,Leo at the beginning of the main-sequence (MS). The evolutionary tracks of \citet{siess2000}/\citet{baraffe2015} predict stellar radii of 0.4/0.5\,\rsun\ for \aum\ at the age of AD\,Leo \citep[i.e., $\sim$100\,Myr according to the stellar evolution models, consistent with literature estimates;][]{shkolnik2009,tuomi2018}. Assuming the conservation of stellar flux and angular momentum during the PMS phase, we predict average magnetic fields of 1.75/0.95\,kG and rotation periods of 1.3/2.4\,d for \aum\ at $\sim$100\,Myr. While the expected rotation period for \aum\ in the case of the \citet{baraffe2015} models (yielding a mass estimate for AD~Leo in better agreement with observations) is comparable to that of AD\,Leo, the predicted field strengths are one order of magnitude larger, demonstrating that magnetic flux conservation \citep[more or less verified in the case of fossil fields, e.g.,][]{landstreet2007} does not apply in the case of dynamos where field amplification is mainly fueled from the kinetic energy reservoir of the convective zone. Both stars are predicted fully- and partly- convective by \citet{siess2000} and \citet{baraffe2015} models, respectively (see Tab.~\ref{tab:star_mag_prop}). While \aum\ will definitely be partly-convective when reaching the MS, AD\,Leo lies slightly above the full-convection threshold of $\sim$0.35\,\msun\ \citep{baraffe1998,baraffe2015}, in the region of \pr-Mass diagram where the magnetic properties of \mdw s are known to significantly differ from those of their more massive counterparts  \citep{donati2008,morin2008,morin2010}.

 
 Similarly, we may wonder what \aum\ looked like at the time its inner accretion disk was exhausted, i.e., at a younger age of under 10~Myr. In particular, did \aum~b have enough time to migrate within the magnetospheric cavity of its host star? In this case, the close-in planet would stand as a signpost for the location of the inner accretion disk and flag where the corotation radius lay at the time of the disk dissipation \citep{lin1996}. Assuming the conservation of angular momentum, this would imply that the disc dissipation occured when \aum\ has a radius of \rs\,$\simeq$\,1\,\rsun, i.e., at a age of about 8\,Myr \citep[according to][]{baraffe2015}. However, the dipole field that the young \aum\ would have needed to carve a magnetospheric cavity as large as 0.066\,au is significantly larger (at least 1.8~kG) than the one it now hosts, even for accretion rates as low as $10^{-10}$\,\msun/yr \citep{bessolaz2008}. Another option is that \aum~b did not have enough time to migrate within the magnetospheric gap before the disc disappeared, or was trapped on its way by interactions with other planets in the system.




This study confirms the ability of SPIRou to carry out precise RV and spectropolarimetric measurements of bright nearby \mdw s. As already demonstrated with simulations, stars with rotation periods of a few days hosting surface features evolving on significantly longer time-scales are excellent targets for high-precision RV follow-ups \citep{klein2020}. In the specific case of low-mass PMS stars like \aum, observing in the nIR rather than in the visible turns out to be a definite advantage to disentangle planet signatures from stellar activity RV signals, the latter being much weaker in this domain. These results are promising for future SPIRou observations of young stars hosting transiting planets to be observed as part of the SPIRou Legacy Survey \citep[e.g.,][]{david2016,david2019,david2019b}.

\aum~b is the first close-in Neptune-sized planet younger than 25\,Myr to have its inner density measured at $>$3$\sigma$. Given the corotation radius of the star (located at 13\,\rs\,=\,0.044\,au), the magnetic field lines are expected to be mostly open under the effect of centrifugal forces at the distance of the planet ($\sim$19~\rs\,=\,0.066~au), implying that magnetic star-planet interactions should operate in the super-alfvenic regime \citep{strugarek2015}. Magneto-hydrodynamical simulations are required to validate this conclusion and further constrain the extended magnetosphere and wind of the star, as well as its interaction with the close-in planet \citep[e.g.,][]{vidotto2017a,vidotto2017b}. We also note that the tilted large-scale magnetic field that we find to be present at the surface of \aum\ may induce a time-dependent wind \citep{vidotto2014}, potentially responsible for the fast moving features identified in the debris disk, as discussed in \citet{Wisniewski2019}.


Given the brightness of its host star, this planet is a primary candidate for an atmospheric characterization with ground-based nIR spectrometers and upcoming space missions like the JWST. Additional more extensive spectropolarimetric observations of \aum\ with SPIRou, with as much as $\sim$100 visits, are crucially needed to accurately pin down the planet mass. These observations will also allow to bring a more precise measurement of the ellipticity of its orbit, which should help retracing the planet formation history and possible dynamical interactions with other putative planets in the system \citep{chatterjee2008,juric2008}. Moreover, given its low mass, \aum~b is unlikely to be responsible for the ejection of the fast moving features in the debris debris \citepalias[see the discussion in][]{plavchan2020}. Hence the interest in conducting more observation campaigns in order to search for further massive bodies whose past orbital evolution could explain the dynamical properties of these features. These observations will also allow us to confirm the surprisingly high DR found in our analysis and investigate the variability of the stellar activity and magnetic field. Finally, an even longer-term spectropolarimetric monitoring of \aum\ would allow to investigate how the fractions of axisymmetric and poloidal magnetic energies and the DR parameters evolve on time-scales of 5-10 years. These parameters are reliable proxies of solar-like magnetic cycles (see Lehmann et al., in prep.) that will help confirming the 5\,yr-activity cycle reported in \citet{ibanez_bustos2018}, thought to be powered by a solar-like $\alpha \Omega$-dynamo process.

\section*{Acknowledgements}
This work is based on observations obtained at the Canada-France-Hawaii Telescope (CFHT) which is operated from the summit of Maunakea by the National Research Council of Canada, the Institut National des Sciences de l'Univers of the Centre National de la Recherche Scientifique of France, and the University of Hawaii. The observations at the Canada-France-Hawaii Telescope were performed with care and respect from the summit of Maunakea which is a significant cultural and historic site. The observations analysed in this work were obtained with SPIRou, an international project led by Institut de Recherche en Astrophysique et Planétologie, Toulouse, France. 

This project was funded by the European Research Council (ERC) under the H2020 research \& innovation programme (grant agreements \#740651 NewWorlds). We acknowledge funding from Agence Nationale de la Recherche (ANR) under contracts ANR-18-CE31-0019 (SPlaSH) and ANR-15-IDEX-02 (Origin of Life at Université Grenobles Alpes).

We warmly thank the referee for valuable comments and suggestions which helped us improving an earlier version of the manuscript.

\section*{Data availability}

The data underlying this article will be available in the Canadian Astronomy Data Center\footnote{\url{https://www.cadc-ccda.hia-iha.nrc-cnrc.gc.ca/}} from 2021/02/28.





\bibliographystyle{mnras}
\bibliography{Bibliography}



\begin{appendix}

\section{Full journal of observations}\label{app:A}

In addition to the journal of observation given in Tab.~\ref{tab:list_obs}, we provide our measurements of RVs for the three methods presented in Sec.~\ref{sec:sec3_1} and the time series of activity indicators described in Sec.~\ref{sec:act_ind} in Tab.~\ref{tab:App_A}.

\begin{table*}
    \centering
    \caption{Information on the spectropolarimetric observations of \aum\ with SPIRou. The first two columns indicate respectively the UT date and the BJD at mid-exposure of each observation. We then list the RVs measured by modeling each Stokes $I$ LSD profile by Gaussian function (reference RVs, column~3), computing the median bisector of each line (column~4), and fitting the first derivative of a Gaussian function to the median-subtracted line profiles (column~5; see Sec.~\ref{sec:sec3_1} for further details about the RV measurement processes). The typical 1$\sigma$ error bar on the RVs is 5\,\ms. The estimated RV photon noise $\sigma_{\rm{ph}}$ is listed in column~6. In columns~7 and~8, we give the values of the longitudinal field, \bl, and the velocity span $V_{\rm{S}}$, computed using the method described in Sec.~\ref{sec:sec3_1} and featuring an error bar of 15\,\ms. Column~9 gives the FWHM of the Stokes $I$ LSD profiles computed from the mask of atomic lines of Land\'e factors $>$1.5 (see Sec.~\ref{sec:act_ind}). The last 3 columns respectively give the indicators based on He\,I, Pa$\beta$ and Br$\gamma$ lines with $\pm$1$\sigma$ error bars (see Sec.~\ref{sec:act_ind}). The flux variation of each line is given by (1-$\gamma$)$\times$ $\delta l$, where $\gamma$ is the activity index listed below, and $\delta l$ us the spectral domain on which the index is computed, equal to 0.28, 0.18 and 0.4\,nm, for He\,I, Pa$\beta$, and Br$\gamma$, respectively. The equivalent widths of the median He\,I, Pa$\beta$ and Br$\gamma$ lines are respectively 0.006\,nm, 0.014\,nm, and $\sim$0.001\,nm.}
    \label{tab:App_A}
\begin{tabular}{ccccccr@{$\pm$}lcr@{$\pm$}lr@{$\pm$}lr@{$\pm$}lr@{$\pm$}l}
\hline
UT Date & BJD  & RV (Ref)  & RV (BIS) & RV (LSD) & $\sigma_{\rm{ph}}$ &  \multicolumn2c{\bl} & $V_{\rm{S}}$ &  \multicolumn2c{FWHM} &  \multicolumn2c{He\,I} & \multicolumn2c{Pa$\beta$} & \multicolumn2c{Br$\gamma$} \\

 &  [2457000+]   & [\ms] & [\ms] & [\ms] & [\ms] & \multicolumn2c{[G]} & [\ms] & \multicolumn2c{[\kms]} & \multicolumn2c{ } & \multicolumn2c{ } & \multicolumn2c{ } \\
\hline
Sep/18 & 1744.8212 & 59.5 & 48.0 & 50.1 & 2.1 & -43.9\, & \,3.1 & -63.1 & 0.38\, & \,0.20 & 1.001\, & \,0.005 & 1.000\, & \,0.004 & 0.993\, & \,0.010 \\
Sep/24 & 1750.7542 & -18.2 & -12.0 & -14.9 & 2.1 & 40.6\, & \,3.7 & 0.0 & 0.00\, & \,0.32 & 1.007\, & \,0.005 & 1.018\, & \,0.005 & 0.992\, & \,0.013 \\
Sep/25 & 1751.7453 & -52.4 & -66.1 & -74.8 & 2.0 & 31.3\, & \,3.2 & 65.1 & -0.74\, & \,0.12 & 1.037\, & \,0.004 & 0.991\, & \,0.004 & 1.003\, & \,0.011 \\
Sep/26 & 1752.7898 & 27.9 & 24.0 & 21.9 & 2.1 & 48.6\, & \,2.9 & -31.0 & -1.36\, & \,0.21 & 1.006\, & \,0.005 & 0.991\, & \,0.004 & 0.994\, & \,0.010 \\
Oct/2 & 1758.7288 & 51.3 & 41.0 & 46.5 & 2.2 & 16.2\, & \,2.8 & -30.0 & -0.01\, & \,0.14 & 0.969\, & \,0.005 & 0.996\, & \,0.004 & 0.993\, & \,0.008 \\
Oct/3 & 1759.8053 & 69.2 & 59.6 & 69.8 & 2.5 & 15.9\, & \,3.2 & -18.0 & 0.79\, & \,0.13 & 1.028\, & \,0.006 & 1.013\, & \,0.004 & 0.995\, & \,0.006 \\
Oct/4 & 1760.7278 & -29.6 & 6.0 & 15.2 & 2.8 & 48.5\, & \,3.7 & 43.0 & -0.14\, & \,0.26 & 1.024\, & \,0.007 & 1.004\, & \,0.005 & 1.006\, & \,0.008 \\
Oct/5 & 1761.7305 & -87.2 & -90.1 & -92.5 & 2.1 & 24.8\, & \,2.9 & 19.0 & -1.04\, & \,0.17 & 1.037\, & \,0.008 & 1.006\, & \,0.005 & 0.997\, & \,0.010 \\
Oct/6 & 1762.7315 & -2.9 & 0.0 & 2.8 & 2.1 & 52.6\, & \,2.7 & -45.0 & 0.11\, & \,0.14 & 1.028\, & \,0.006 & 1.004\, & \,0.004 & 1.007\, & \,0.010 \\
Oct/8 & 1764.7571 & 34.5 & 48.0 & 62.7 & 2.1 & 23.6\, & \,3.9 & 21.0 & 1.64\, & \,0.25 & 1.017\, & \,0.005 & 1.012\, & \,0.004 & 1.014\, & \,0.013 \\
Oct/9 & 1765.7694 & -39.8 & -36.0 & -35.1 & 2.0 & 34.4\, & \,3.2 & 44.0 & 0.25\, & \,0.22 & 1.020\, & \,0.005 & 1.012\, & \,0.004 & 1.014\, & \,0.010 \\
Oct/13 & 1769.7438 & 22.1 & 48.0 & 55.0 & 2.0 & 32.6\, & \,3.1 & 5.0 & 0.88\, & \,0.27 & 0.998\, & \,0.006 & 1.006\, & \,0.004 & 1.013\, & \,0.009 \\
Oct/14 & 1770.7407 & -41.6 & -42.0 & -33.6 & 2.2 & 33.4\, & \,3.4 & 16.0 & -0.32\, & \,0.17 & 1.012\, & \,0.004 & 1.000\, & \,0.003 & 1.000\, & \,0.008 \\
Oct/15 & 1771.7212 & -55.8 & -30.0 & -36.2 & 2.3 & 16.4\, & \,3.1 & 61.1 & -0.89\, & \,0.19 & 1.016\, & \,0.005 & 1.000\, & \,0.003 & 1.013\, & \,0.007 \\
Oct/16 & 1772.7416 & 19.4 & 24.0 & 31.1 & 2.4 & 47.2\, & \,3.4 & -103.1 & 0.48\, & \,0.31 & 0.984\, & \,0.005 & 0.995\, & \,0.003 & 0.983\, & \,0.009 \\
Oct/31 & 1787.7155 & 28.3 & 6.0 & 4.1 & 2.5 & -4.7\, & \,3.2 & -57.1 & 1.55\, & \,0.19 & 0.971\, & \,0.005 & 0.979\, & \,0.004 & 0.999\, & \,0.007 \\
Oct/32 & 1788.7045 & 41.3 & 60.1 & 58.5 & 2.3 & 4.2\, & \,3.3 & -96.1 & 1.16\, & \,0.20 & 0.986\, & \,0.005 & 0.992\, & \,0.004 & 0.990\, & \,0.006 \\
Nov/2 & 1789.7367 & -32.6 & -24.0 & -22.3 & 2.1 & 82.5\, & \,3.2 & 43.0 & -0.80\, & \,0.19 & 0.991\, & \,0.004 & 0.997\, & \,0.004 & 1.000\, & \,0.007 \\
Nov/3 & 1790.7010 & -90.1 & -90.1 & -85.0 & 2.1 & 12.4\, & \,3.6 & 75.1 & -1.46\, & \,0.15 & 1.009\, & \,0.005 & 1.006\, & \,0.004 & 1.012\, & \,0.012 \\
Nov/4 & 1791.6983 & 2.1 & 0.0 & 5.6 & 2.1 & 40.3\, & \,2.8 & 31.0 & -0.67\, & \,0.14 & 0.989\, & \,0.005 & 0.992\, & \,0.004 & 0.994\, & \,0.007 \\
Nov/5 & 1792.6976 & 19.1 & 18.0 & 18.8 & 2.0 & -24.3\, & \,3.1 & -30.0 & 0.42\, & \,0.16 & 0.976\, & \,0.003 & 0.975\, & \,0.004 & 0.988\, & \,0.007 \\
Nov/9 & 1796.6859 & -13.5 & -24.0 & -24.3 & 2.7 & 43.7\, & \,3.6 & -56.6 & 0.18\, & \,0.15 & 0.974\, & \,0.007 & 1.001\, & \,0.004 & 1.000\, & \,0.008 \\
Nov/10 & 1797.7098 & 19.7 & 37.5 & 56.6 & 2.7 & -48.0\, & \,3.7 & 27.0 & -0.07\, & \,0.35 & 0.969\, & \,0.007 & 0.987\, & \,0.007 & 0.989\, & \,0.011 \\
Nov/11 & 1798.6873 & 27.0 & 36.0 & 38.2 & 2.5 & 69.5\, & \,3.5 & -47.0 & 1.71\, & \,0.10 & 0.978\, & \,0.004 & 0.995\, & \,0.004 & 1.007\, & \,0.005 \\
Nov/12 & 1799.6883 & -10.3 & -30.0 & -30.3 & 2.7 & 70.0\, & \,3.9 & -28.0 & 0.60\, & \,0.24 & 0.998\, & \,0.006 & 1.010\, & \,0.006 & 0.999\, & \,0.009 \\
Nov/13 & 1800.6896 & -46.2 & -42.0 & -56.5 & 2.1 & 8.8\, & \,3.0 & 34.0 & -0.66\, & \,0.24 & 1.001\, & \,0.006 & 1.012\, & \,0.005 & 1.001\, & \,0.009 \\
Nov/14 & 1801.6873 & 0.0 & -12.0 & -12.2 & 2.2 & 44.3\, & \,3.2 & -33.0 & -0.13\, & \,0.18 & 0.997\, & \,0.006 & 1.002\, & \,0.006 & 1.009\, & \,0.011 \\
\hline
\end{tabular}
\end{table*}

\section{Simulating the modeling of our RV time-series including excess uncorrelated noise}\label{app:B}

\begin{figure}
    \centering
    \includegraphics[width=\linewidth]{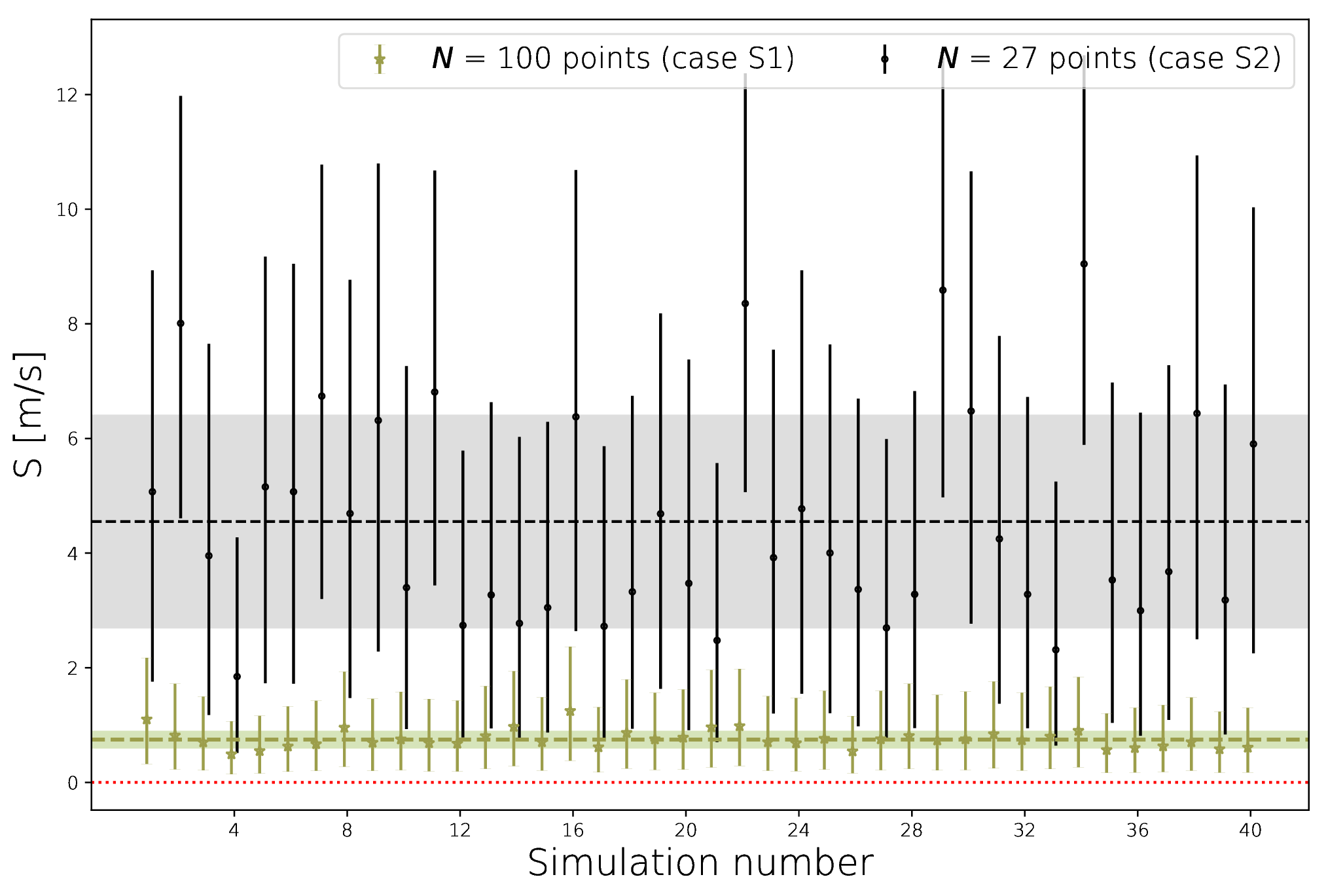}
    \caption{Best estimates of $S$ obtained when modeling 40 simulated RV times series, containing either 27 points (case S2, black dots) or 100 points (case S1, green stars), using the estimation process described in Sec.~\ref{sec:sec3_2}. The value of $S$ to recover (i.e., 0.0\,\ms) is indicated by the red dotted line, whereas the mean values (resp. standard deviations) of $S$ recovered in cases S1 and S2 are respectively indicated by green and black dashed lines (resp. light green and gray bands). For clarity, the $S$ values obtained in case S1 are slightly shifted horizontally from those obtained in case S2.}
    \label{fig:simu_s4}
\end{figure}

We built a synthetic RV curve containing the GP prediction of the stellar activity signal (in the case of the reference RVs, i.e., line 1 of Tab.~\ref{tab:results_rv}) and a planet signature of semi-amplitude 8.5\,\ms. From this curve, we created two different data sets, called S1 and S2, containing respectively $N$\,=\,100 evenly-sampled data points and $N$\,=\,27 data points at the same epochs as our observations. In both cases, we added a random noise including a photon noise of 2\,\ms\ RMS, an additional uncorrelated Gaussian noise of 4\,\ms\ RMS, and finally set the 1$\sigma$ RV uncertainties on each synthetic data point to 5\,\ms. We then ran our estimation process, including a fit of the excess uncorrelated noise $S$ (see Sec.~\ref{sec:sec3.3}), on 40 of these simulated data sets, each with a different noise realization. Ideally, one would expect $S$ to be null, given that the assumed formal error bars already account for the injected noise.

The resulting distribution of $S$ is shown in Fig.~\ref{fig:simu_s4}. Whereas no significant excess noise is recovered in case S1 (mean value of 0.7\,$\pm$\,0.2\,\ms), $S$ is systematically over-estimated in case S2, with a mean recovered value of 4.5\,$\pm$\,2.0\,\ms. This suggests that the value of $S$ we derived from our data, found to be 6.0$^{+3.8}_{-3.1}$\,\ms, i.e., consistent with the average value of 4.5\,$\pm$\,2.0\,\ms\ recovered in our simulations, is most likely over-estimated.

\section{Modeling activity indicators}\label{app:C}

{\renewcommand{\arraystretch}{1.25} 
\begin{table*}
    \centering
    \caption{Results of the fit to the time series of the activity indicators analysed in Sec.~\ref{sec:act_ind}. The first two lines are respectively the typical formal uncertainty in the time series ($\bar{\sigma}$), and the excess of uncorrelated noise recovered by the model. Lines~3 to~7 list the best estimates of the four hyperparameters (written in bold when fixed in the MCMC process), and the constant offset. Finally, we give the \crr\ of the MCMC fit in line~7.}
    \label{tab:fit_indic}
    \begin{tabular}{cccccccc}
        \hline
     Indicator   & RVs  & \bl   &  $V_{\rm{S}}$ & FWHM & He\,I & Pa$\beta$\\
        \hline
        $\bar{\sigma}$ & 5\,\ms\ &  3.2\,G & 15\,\ms  & 0.19\,\kms & 0.005 & 0.004  \\
        $S$ & 6.0$^{+3.8}_{-3.1}$\,\ms\ & 4.4$^{+2.7}_{-2.4}$\,G & 16$^{+13}_{-10}$\,\ms & 0.55$^{+0.21}_{-0.19}$\,\kms  &  0.010$^{+0.003}_{-0.002}$ & 0.006\,$\pm$\,0.002  \\
       $\theta_{1}$ & 43$^{+11}_{-8}$\,\ms &  68$^{+40}_{-27}$\,G & 36$^{+11}_{-8}$\,\ms & 0.9$^{+0.4}_{-0.3}$\,\kms & 0.03\,$\pm$\,0.01 & 0.011$^{+0.006}_{-0.004}$  \\
       $\theta_{2}$ [d] &  \textbf{100} & 105$^{+67}_{-60}$ & \textbf{100} & \textbf{100} & \textbf{100} & \textbf{100}\\
       $\theta_{3}$ [d] & 4.84\,$\pm$\,0.01  & 4.83\,$\pm$\,0.02 & 4.89$^{+0.04}_{-0.02}$ & 4.84\,$\pm$\,0.04 & 4.87$^{+0.06}_{-0.05}$ & 4.99$^{+0.11}_{-0.09}$  \\  
      $\theta_{4}$ & \textbf{0.4} & 0.7$^{+0.4}_{-0.2}$ & 0.36$^{+0.17}_{-0.12}$  &  \textbf{1.0} & \textbf{1.0}  & \textbf{1.0} \\       
      Constant offset & 0$^{+19}_{-18}$\,\ms & 44$^{+61}_{-39}$\,G & 4\,$\pm$\,26\,\ms & 0.0\,$\pm$\,0.6\,\kms & 1.00\,$\pm$\,0.02 & 1.00\,$\pm$\,0.01 \\
       \crr  & 0.8  & 0.8 & 4.0 & 7.8 & 3.2 & 2.4 \\  
        \hline
    \end{tabular}
\end{table*}}

\begin{figure*}
    \centering
    \includegraphics[width=\linewidth]{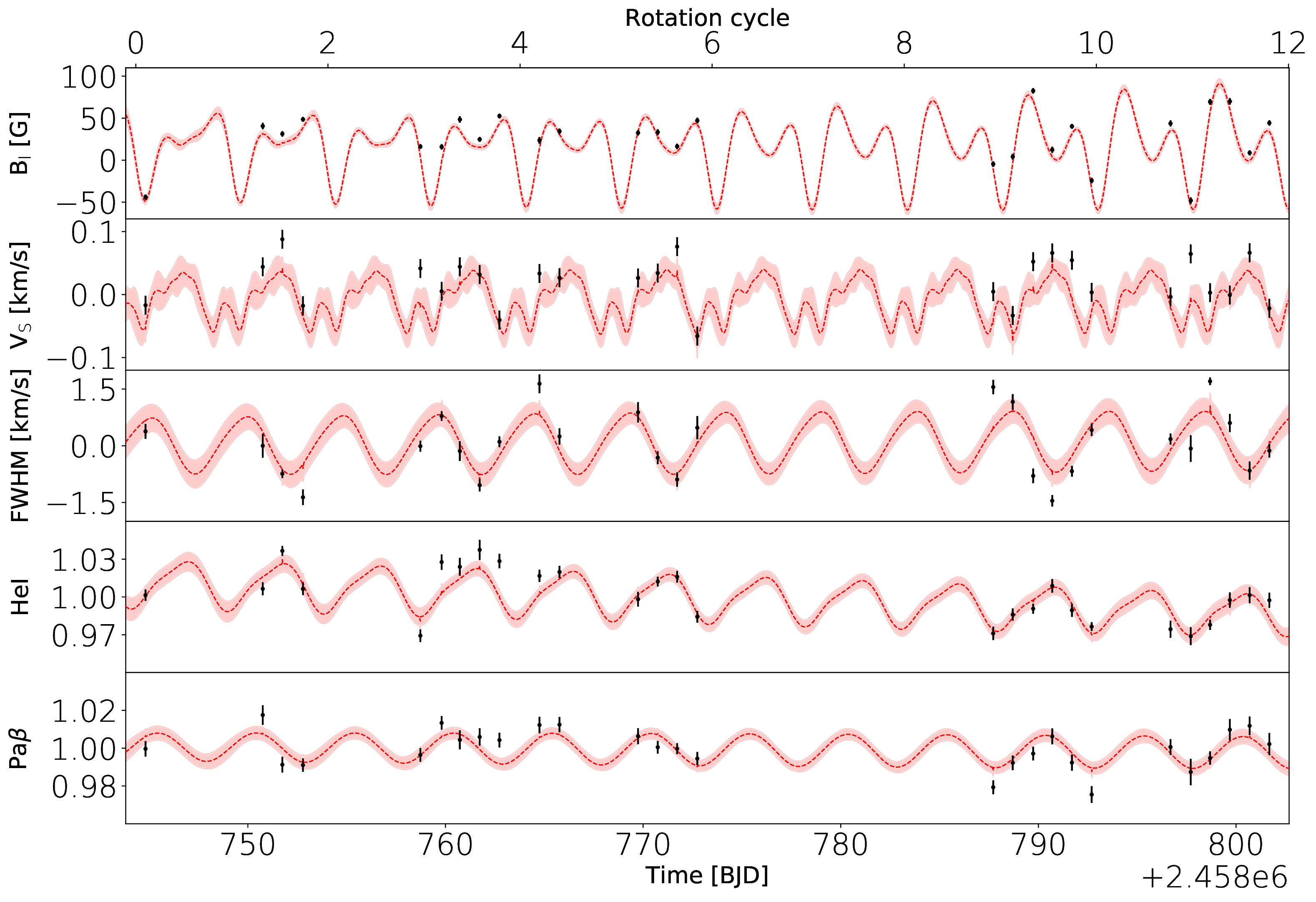}
    \caption{Best GP predictions (red dotted lines with $\pm$1$\sigma$ error bands) of the time series of each indicator analysed in Sec.~\ref{sec:act_ind} (black dots). The estimation of the hyperparameters of the GP (i.e., \hypv) is carried out following the method described in Sec.~\ref{sec:sec3_2} and adopting the prior densities detailed in Sec.~\ref{sec:act_ind}.}
    \label{fig:pred_GP_ind}
\end{figure*}

\begin{figure*}
    \centering
    \includegraphics[width=\linewidth]{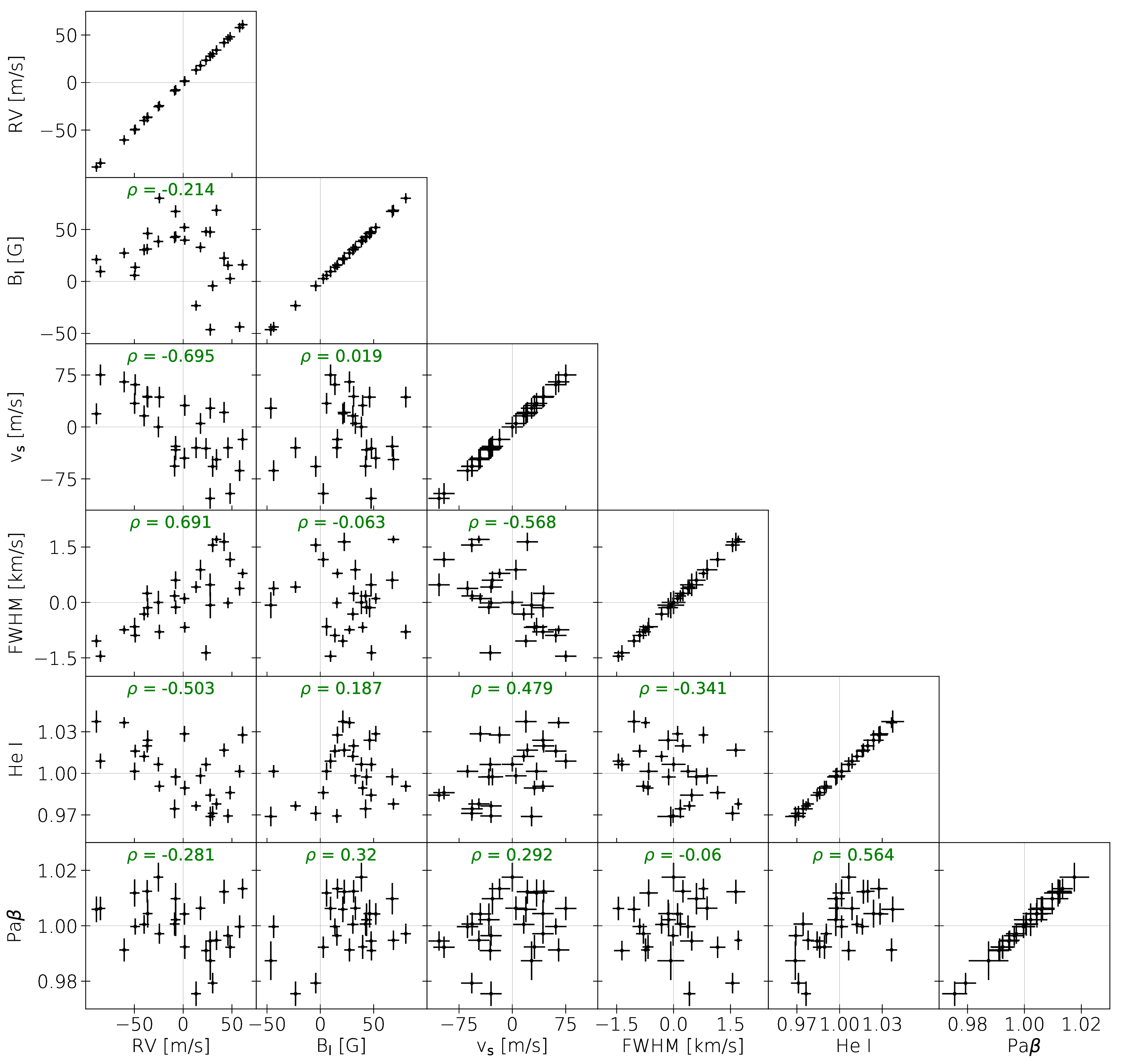}
    \caption{Correlation plots of the of activity indicators analysed in Sec.~\ref{sec:act_ind}. The Pearson correlation coefficient, $\rho$, is written in green at the top of each panel.}
    \label{fig:ind_correl_all}
\end{figure*}

To complement the analysis of stellar activity indicators of Sec.~\ref{sec:act_ind}, we give results of the GP fit to the time-series of \bl, $V_{\rm{S}}$, FWHM, He\,I, Pa$\beta$ in Tab.~\ref{tab:fit_indic}, and show the best prediction in Fig.~\ref{fig:pred_GP_ind}. Br$\gamma$, that does not appear rotationally-modulated, is not displayed here. We also plot each pair of activity indicator and give the associated Pearson correlation coefficient in Fig.~\ref{fig:ind_correl_all}.

\end{appendix}

\bsp	
\label{lastpage}
\end{document}